\documentclass[a4paper,11pt]{article}

\usepackage{jcappub}
\usepackage{subcaption}  

\usepackage{graphicx}
\usepackage{tikz}

\title{Quantum corrections to symmetron fifth forces for planar sources}
\author{Peter Millington,}
\author{Michael Udemba}
\affiliation{Department of Physics and Astronomy, University of Manchester, Manchester M13 9PL, U.K.}

\emailAdd{michael.udemba@manchester.ac.uk}
\emailAdd{peter.millington@manchester.ac.uk}

\abstract{We provide a semi-analytic calculation of the first quantum corrections to the symmetron fifth force around a planar source with nonzero thickness. We find a suppression of the fifth force compared with the classical prediction within a Compton wavelength of the source, which is of order 10\% in the parameter region relevant to experiments like CANNEX, while the fifth force is enhanced at larger distances from the source. The resulting change in the spatial profile of the fifth force may be relevant to current and near future terrestrial and astrophysical tests of force laws, and has implications for the optimisation of experimental geometries, including atom interferometers. This work provides a key benchmark for future numerical studies of quantum-corrected fifth forces in screened scalar-tensor theories of gravity.}


\begin{document}
\tikzset{->-/.style={decoration={
  markings,
  mark=at position .5 with {\arrow{>}}}, postaction={decorate}}}

\maketitle
\section{Introduction}
\label{sec_1}
One popular solution to the open problems of cosmology is to introduce new scalar degrees of freedom. They may account for, among other things, the accelerated expansion of the universe \cite{Mehta2026, Kepuladze2026, Beesham2026} and flat galaxy rotation curves \cite{Almurr2026, Zaregonbadi2025, Dabbaghsaz2025}. One generally expects new light scalars to mediate long-range fifth-forces between matter particles. This can pose a theoretical challenge, since Solar-System and near-Earth tests of gravity have yet to yield a deviation from known physics \cite{Feleppa2026}. Furthermore, recent results from the Atacama Cosmology Telescope strongly suggest that gravity obeys the inverse-square law over very large (\(\gtrsim\) Mpc) distances \cite{Gallardo2026}, while theories of modified gravity which introduce new scalar fields generically predict otherwise \cite{Will2001, Clifton2012}.

Nonlinear scalar-tensor theories may provide a means of modifying gravity in a way that is consistent with astrophysical and cosmological observations. Their nonlinear equations of motion allow for the properties of the fifth force to depend on the background environment. In particular, their model parameters can be chosen such that the fifth force is negligible in environments like the Solar System. In models which exhibit Vainshtein screening, this is achieved through higher-derivative terms in the Lagrangian \cite{Amendola2025}. Models like the chameleon have an effective mass (equivalently, fifth-force range) which depends on the background matter density \cite{Zaregonbadi2025}. For symmetron \cite{Hinterbichler:2011ca} and symmetron-like \cite{Kading:2023hdb} models, the coupling of the fifth force to matter is background dependent, and is suppressed in environments at least as dense as the model's critical density \cite{BachsEsteban2025}.

Many of the strongest constraints on new long-range force carriers come from high-precision tabletop experiments. For the symmetron, constraints in the meV mass range are found primarily through interferometry \cite{Burrage2016-aq}, measurements of the magnetic moment of the electron \cite{Brax2022} and torsion balance experiments \cite{Sun:2024qis, Krishak:2020opb, Wagner:2012ui}. Casimir experiments like CANNEX probe the eV range \cite{Sedmik2021, Fischer:2024gni}. The most sweeping constraints are derived from hydrogen and muonium spectroscopy \cite{PhysRevD.107.044008}, neutron interferometry \cite{Dvorak:2026avo, Fischer:2023eww} and the qBounce experiment \cite{Jenke:2020obe, Jenke:2019qkw}, ranging meV through to GeV mass scales. Recently, searches using levitated sensors have also emerged \cite{Yin:2025uzf, Li:2024ynr}.

However, these constraints and much of the literature on scalar-tensor theories, nonlinear or otherwise, is based on analysis at the level of the classical field equations.  A growing body of work is emerging in which the quantum corrections to nonlinear scalar-tensor theories are being accounted for. Recent analyses have suggested that the constraints on the symmetron and chameleon \cite{Brax2019, BRAX2023101294} models in particular may be significantly modified by loop corrections.

In a previous work \cite{Millington2026}, we presented an analytic approximation of the loop corrections for the symmetron field, in the limit of large spherical sources. Our computations suggested that the quantum-corrected symmetron force may be significantly weaker than the classical (or tree-level) prediction. The magnitude of this correction scaled almost linearly with the self-interaction and was essentially constant in the field mass. A key assumption which allowed us to greatly simplify our derivation was to take the source radius much larger than the Compton wavelength of the field. This allowed us to employ an analogy of Coleman's thin-wall approximation, as usually applied to calculate decay rates between quasi-degenerate vacuum states~\cite{Coleman1977, Coleman1977e, Callan1977}. Doing so meant the source was effectively infinite in thickness. This work is the second in this series of papers, providing analytic and semi-analytic calculations that will provide a benchmark for future numerical calculations of the quantum corrections to fifth forces. We build upon our previous results by considering a planar source of finite thickness. To keep the semi-analytic calculation tractable, we will also take the source to be wide enough such that boundary effects may be neglected. We find a similar reduction in the strength of the fifth force, relative to the classical prediction. within one Compton wavelength of the source, but an enhancement of the fifth force beyond this, again in accord with the results of ref.~\cite{Millington2026}.

The presentation of this paper is as follows. In the following section, we give the derivation of the symmetron field profile in the vicinity of a planar source of finite thickness and infinite extent. It happens to be identical to the symmetron field in one spatial dimension. Then, in Section 3, we perform a spectral analysis of the field's fluctuation operator. As was the case in our previous work, this section serves a dual purpose. First, it provides a novel verification of the stability of the classical configuration. Second, it prepares the calculation of the Green's function. Next, in Section 4, we outline the derivation of the Green's function. Then, in Section 5, we present the calculation of the tadpole contribution, the one-loop field profile and the spatial variation of the one-loop force. As was also identified in our previous paper, we observe a reduction in the fifth force near the surface of the source and a local increase a few Compton wavelengths away. Unlike in the planar limit of the spherical problem, we will be unable to express the coincident Green's function as a sum of two terms in which only one contains the UV divergence. Consequently, much of the calculations in this section are performed numerically. Finally, we discuss our conclusions and directions for future work in Section 6.

In Sections 4 and 5, we make use of results from the analysis of the Heun and Lam{\'e} equation. We outline useful results regarding these functions in Appendix A. In particular, we demonstrate that, while the fluctuations may equivalently be expressed as Lam{\'e} functions, their representation as Heun functions is far more practical for our purposes. In addition, the technical manipulations required to obtain the results for the Green's function in Section 4 are given in Appendix B.
\section{Classical Configuration}
\label{sec_2}
The one-loop field \(\phi_\text{qu}\) may be expressed in terms of the tree-level configuration \(\phi_\text{cl}\) as \(\phi_\text{qu} = \phi_\text{cl} + \delta\phi\). The quantum correction \(\delta\phi\) is given by \cite{Millington2026, Garbrecht:2015oea}
\begin{equation}
    \label{eq:q_correction}
    \delta\phi(x) = \int\text{d}^4y\,G(x, y)[\phi_\text{cl}]\Pi(y)[\phi_\text{cl}] \phi_\text{cl}(y)\:,
\end{equation}
where \(G\) is the Green's function and \(\Pi(y)[\phi_\text{cl}]=G(y,y)\) is the one-loop tadpole. Note that the Green's function appearing explicitly and in the tadpole must be calculated in the background of the classical field $\phi_\text{cl}$, thus accounting fully for the breaking of translational invariance by the presence of the source and the resulting spatially varying classical field configuration. This is the central challenge of computing the first quantum corrections to the fifth force. To this end, we first need to compute the classical field configuration.

In this section, we provide a brief overview of the symmetron model. In addition, we summarise results previously obtained by Burrage et al. \cite{PhysRevD.99.024045}, Brax and Pitschmann \cite{PhysRevD.97.064015, PhysRevD.103.084013}. Their analyses considered one-dimensional sources of finite extent. Given a matter distribution in the shape of a plane, which is large enough to allow us to neglect distortions near the boundary, the surrounding symmetron field is effectively one dimensional.

The symmetron is a scalar field theory characterised by the action \cite{Hinterbichler:2011ca}
\begin{equation}
    S[\phi] = \int\text{d}^4x\,\sqrt{-g}\left[\frac{M_\text{Pl}^2}{2}R + \frac{1}{2}g^{\mu\nu}\partial_\mu\phi\partial_\nu\phi-V(\phi)\right] + \int\text{d}^4x\,\sqrt{-\tilde{g}}\mathcal L_\text{m}(\tilde{g}_{\mu\nu})\:,
\end{equation}
where \(g\) is the determinant of the Einstein-frame metric \(g_{\mu\nu}\), \(R\) is the Ricci scalar, \(M_\text{Pl}\) is the reduced Planck mass and \(\mathcal L_\text{m}\) denotes the Lagrangian density of matter fields. These fields are minimally coupled to the Jordan-frame metric,
\begin{equation}
    \tilde{g}_{\mu\nu} = A(\phi)^2g_{\mu\nu}\:,
\end{equation}
which is related to the Einstein-frame metric by a positive rescaling \(A(\phi)^2\). By the stationary action principle, one obtains the equation of motion of the scalar field \(\phi\),
\begin{equation}
    \square\phi + \frac{\text{d}V}{\text{d}\phi} - A(\phi)^3\frac{\text{d}A}{\text{d}\phi}\,\tilde T = \square \phi + \frac{\text{d}V_\text{eff}}{\text{d}\phi} = 0\:,
\end{equation}
\(\tilde T\) is the trace of the matter energy-momentum tensor in the Jordan frame, and \(V_\text{eff}\) is the effective potential. Our matter distribution is non-relativistic and pressureless, so we may write \(\tilde T \approx -\tilde\rho\), with \(\rho = A(\phi)^3\tilde\rho\), as its energy density.

We take the potential for the symmetron field to be \cite{Hinterbichler2010, Burrage2021}
\begin{equation}
    V(\phi) = -\frac{1}{2}\mu^2\phi^2 + \frac{1}{4}\lambda\phi^4 + \frac{\mu^4}{4\lambda}\:,
\end{equation}
where \(\mu^2 > 0\) is the mass term and \(\lambda\) is the self-coupling. The additive constant is included for computational convenience and corresponds to setting the energy density of the symmetry breaking vacuum to zero. Its inclusion has no effect on the resulting dynamics. However, it amounts to tuning the symmetron field's contribution to the cosmological constant. Since our focus is fifth forces, we shall make no further comment on this aspect of the symmetron model's phenomenology. It is also convenient to cast the theory in terms of the effective potential
\begin{equation}
    V_\text{eff}(\phi) = V(\phi) + A(\phi)\rho\:,
\end{equation}
where \(A\) is a coupling function and \(\rho\) describes the background matter density. For the symmetron field, we have
\begin{equation}
    A(\phi) = \frac{1}{2M^2}\phi^2\:,
\end{equation}
where \(M\) is a mass scale corresponding to the strength of the symmetron's matter coupling. Then, the effective potential is given by 
\begin{equation}
    V_\text{eff}(\phi) = \frac{1}{2}\left(\frac{\rho}{M^2} - \mu^2\right)\phi^2 + \frac{1}{4}\lambda\phi^4 + \frac{\mu^4}{4\lambda}\:.
\end{equation}
The symmetron screening mechanism, which provides a means for the fifth force to evade local tests of gravity, is realised as follows: (i) in regions of high density, in which the coefficient of the quadratic term in the potential is positive, the expectation value of the field $\phi_{\rm cl}$ is zero, and the classical fifth force $F\propto -\phi_{\rm cl}\nabla \phi_{\rm cl}$ vanishes; (ii) in regions of low density, in which the same coefficient is negative, the field $\phi_{\rm cl}$ acquires a non-zero value and the classical fifth force is non-vanishing.

For simplicity, we consider a scenario in which the system has come to rest in its equilibrium configuration. As such, we seek the configuration which obeys the static equation of motion,
\begin{equation}
    \label{eq:eom}
    \nabla^2\phi = \frac{\text{d}V_\text{eff}}{\text{d}\phi}\:,
\end{equation}
along with the boundary condition \(\phi\rightarrow v = \mu/\sqrt{\lambda}\) at spatial infinity and the requirement that \(\nabla\phi\) vanishes at the origin. Let \(2R\) be the thickness of the matter source. We denote its uniform density by \(\rho_0\). Its density distribution is therefore proportional to the rectangular function,
\begin{equation}
    \rho(x) = \rho_0\operatorname{rect}\left(\frac{x}{2R}\right) \equiv \begin{cases} 0\:,&x<R\\ \rho_0\:, & -R\leq x \leq R\\ 0\:, & x > R\end{cases}\:.
\end{equation}
We reiterate that our modelling assumptions make the field effectively one dimensional, allowing us to swap the gradient operator \(\nabla\) in eq.~\ref{eq:eom} for an ordinary derivative in \(x\). Furthermore, due to the symmetry of the effective potential, the field is an even function.

The static equation may now be written in the form
\begin{equation}
    \label{eq:other_order_reduction}
    \frac{1}{2}\frac{\text{d}}{\text{d}x}\left(\frac{\text{d}\phi}{\text{d}x}\right)^2 = \frac{\text{d}V_\text{eff}}{\text{d}x}\:,
\end{equation}
which can be directly integrated on the domain \((x_1, x_2)\subseteq \mathbb R\) to yield
\begin{equation}
    \left(\left.\frac{\text{d}\phi}{\text{d}x}\right\vert_{x = x_2}\right)^2 - \left(\left.\frac{\text{d}\phi}{\text{d}x}\right\vert_{x = x_1}\right)^2 = 2\left[V_\text{eff}(\phi(x_2)) - V_\text{eff}(\phi(x_1))\right]\:.
\end{equation}
Since the potential is symmetric, we may restrict ourselves to solving on the positive domain. If we assume that the field does not vanish at the origin then the lowest energy configuration will be even under reflection about the plane of the source. Furthermore, in the positive domain, we denote by \(\phi_\text{ext}(x)\) and \(\phi_\text{int}(x)\), respectively, the field exterior and interior to the source. Separately solving for them is simply a matter of picking appropriate values for \(x_1\) and \(x_2\). The full profile, in the positive domain, given by
\begin{equation}
    \phi(x) = \Theta(R - x)\phi_\text{int}(x) + \Theta(x - R)\phi_\text{ext}(x)\:,
\end{equation}
is then uniquely determined by imposing continuity and differentiability (continuity of the first derivative) at \(x = R\). 

For the exterior solution, we choose \(x_1 = x > R\) and \(x_2 = \infty\) to arrive at the equation
\begin{equation}
    \left(\frac{\text{d}\phi_\text{ext}}{\text{d}x}\right)^2 = - \mu^2\phi_\text{ext}^2 + \frac{\lambda}{2}\phi_\text{ext}^4 + \frac{\mu^4}{2\lambda}\:.
\end{equation}
Writing \(\chi_\text{int/ext} = \phi_\text{int/ext} v^{-1}\) and factorising yields the equation
\begin{equation}
    \frac{\text{d}\chi_\text{ext}}{\text{d}x} = \frac{\mu}{\sqrt{2}}\left(1 - \chi_\text{ext}^2\right)\:.
\end{equation}
We take the positive root since \(\chi < 1\) and, for \(x>0\), the field and its first derivative are positive functions. By comparison with the identity
\begin{equation}
    \frac{\text{d}}{\text{d}x}\tanh(x) = \operatorname{sech}^2(x) = 1 - \tanh^2(x)\:,
\end{equation}
it is clear that the solution may be written
\begin{equation}
    \chi_\text{ext}(x) = \tanh\left[\frac{\mu}{\sqrt{2}}\left(x - R\right) + \operatorname{arctanh}\left(\chi_{R}\right)\right]\:,
\end{equation}
where \(\chi_{R} \equiv \chi(R)\). Interior to the source, we choose \(x_1 = 0\) and \(x_2 = x < R\). This yields the equation
\begin{equation}
    \left(\frac{\text{d}\phi_\text{int}}{\text{d}x}\right)^2 = \left(\frac{\rho_0}{M^2} - \mu^2\right)\left(\phi_\text{int}^2 - \phi_0^2\right) + \frac{\lambda}{2}\left(\phi_\text{int}^4 - \phi_0^4\right)\:,
\end{equation}
where \(\phi_0\equiv\phi(0)\). Normalising the field with respect to the vacuum expectation value $v=\mu/\sqrt{\lambda}$ allows to simplify this equation, yielding
\begin{equation}
    \left(\frac{\text{d}\chi_\text{int}}{\text{d}x}\right)^2 = \frac{\mu^2}{2}\left(\chi_\text{int}^2 - \chi_0^2\right)\left(\chi_\text{int}^2 + 2\xi^2 - \chi_0^2\right)\:,
\end{equation}
where we have defined
\begin{equation}
    \xi = \sqrt{\frac{\rho_0}{\mu^2M^2} - 1 + \frac{\phi_0^2}{v^2}}\:.
\end{equation}
It happens to be close to the effective mass of interior scalar fluctuations \(m_-\), in units of \(\mu\). They are related by \(m_-^2 = 2\chi_0^2 + \mu^2\xi^2\). Taking roots of both sides yields the order reduction
\begin{equation}
    \frac{\text{d}\chi_\text{int}}{\text{d}x} = \frac{\mu}{\sqrt{2}}\sqrt{\left(\chi^2 - \chi_0^2\right)\left(\chi^2 + 2\xi^2 - \chi_0^2\right)}\:.
\end{equation}
We now invert the above equation and obtain the integral expression
\begin{equation}
    \mu x = \sqrt 2\int_{\chi_0}^{\chi_\text{int}(x)}\frac{\text{d}\chi}{\sqrt{\left(\chi^2 - \chi_0^2\right)\left(\chi^2 + 2\xi^2 - \chi_0^2\right)}} = \frac{1}{\xi}\int_1^{\chi_\text{int}/\chi_0}\frac{\text{d}t}{\sqrt{(t^2 - 1)((1 - k^2)t^2 + k^2)}}\:,
\end{equation}
where we have used the substitution \(t = \chi_\text{int}/\chi(0)\) and defined the parameter
\begin{equation}
    k^2 = 1 - \frac{\phi_0^2}{2\xi^2 v^2}\:,
\end{equation}
which will turn out to be an elliptic modulus. By comparison with the definition of the Jacobi elliptic function \(\operatorname{nc}\),
\begin{equation}
    \label{eq:cn_integral}
    u = \int_1^{\text{nc}\left(u, m\right)}\frac{\text{d}t}{\sqrt{(t^2 - 1)((1 - m)t^2 + m)}}\:,
\end{equation}
we find that the interior field is given by
\begin{equation}
    \phi_\text{int}(x) = \phi_0\operatorname{nc}(\xi\mu x, k^2)\:.
\end{equation}
We will make use of some properties of Jacobi elliptic functions later on in our analysis. A summary of them is given in Appendix C.

\begin{figure}
    \begin{center}
        \includegraphics{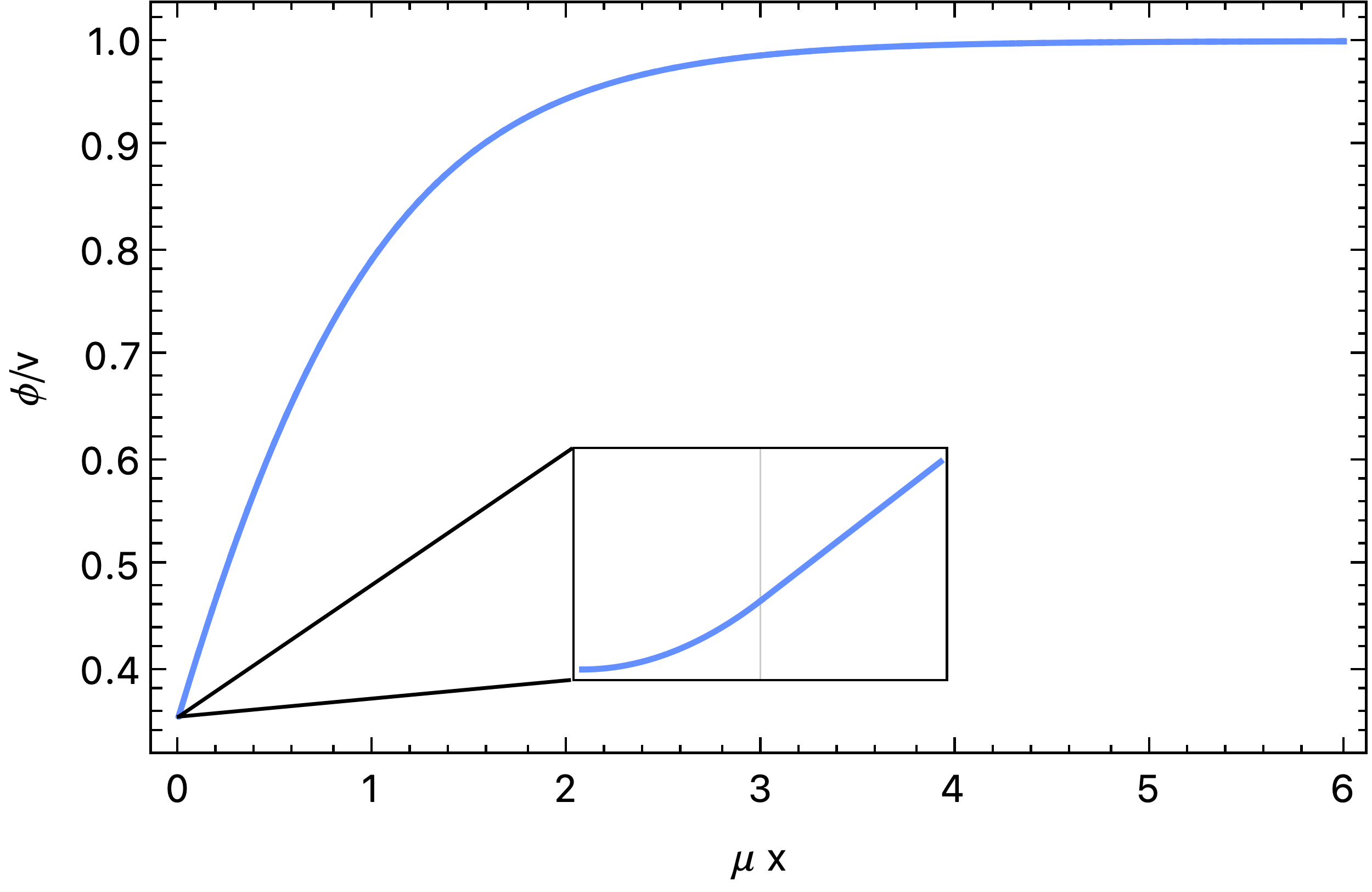}
    \end{center}
    \caption{A plot of the symmetron field from the surface of a planar source. Parameter values correspond to CANNEX \cite{Almasi2015, Sedmik2021}: \(\rho_0 = \rho_{\text{SiO}_2}\), \(\mu = 10^{-1}\) eV, \(M = 10^6\) eV, \(R = 3\,\)mm and \(\lambda = 0.5\) give \(\chi_0\approx 0.35598\). The vertical line on the inset plot is at the surface of the source.}
    \label{fig:1dprofile}
\end{figure}

From continuity at the surface of the source, it follows that
\begin{equation}
    \chi_0\operatorname{nc}(\xi\mu R, k^2) = \chi_{R}\:,
\end{equation}
and so the value of the field at the surface is determined by its value at the origin. By differentiability, we have
\begin{equation}
    \nonumber (\chi_{R}^2 - \chi_0^2)\left(\chi_{R}^2 + 2\xi^2 - \chi_0^2\right) = \left(1 - \chi_{R}^2\right)^2
\end{equation}
    and, by combining the two, we arrive at an equation for the value of the field at the origin $\chi_0$, given by
\begin{equation}
    \Big(\chi_0^2\operatorname{nc}(\xi\mu R, k^2)^2 - \chi_0^2\Big)\Big(\chi_0^2\operatorname{nc}(\xi\mu R, k^2)^2 + 2\xi^2 - \chi_0^2\Big) = \Big(1 - \chi_0^2\operatorname{nc}(\xi\mu R, k^2)^2\Big)^2\:.
\end{equation}
The above equation is transcendental, and thus there is no closed-form expression for \(\chi_0\). It is easy enough, however, to numerically approximate \(\chi_0\) and thus arrive at an estimate for the 1D field profile. A plot is given in Figure.~\ref{fig:1dprofile}, for parameter values relevant to CANNEX \cite{Sedmik2021}. We plot the field profile from the origin of the source, but note that the full profile is given by a reflection about the vertical axis. 

\section{Spectral Analysis of the Fluctuation Operator}
\label{sec_3}
In this section, we provide an analysis of the discrete eigenmodes of the 1+1 dimensional fluctuation operator. This serves two purposes:\ first, this allows to confirm the stability of the classical field configuration, and second, it acts as a warm-up to the calculation of the Green's function of the next section.

The second functional derivative of the action evaluated on the classical configuration \(\phi_\text{cl}\) is
\begin{equation}
    \label{eq:1d_sfd}
    \left.\frac{\delta^2S}{\delta \phi(x')\delta\phi(x)}\right\vert_{\phi = \phi_\text{cl}} = \left(-\frac{\partial^2}{\partial t^2} + \frac{\partial^2}{\partial x^2} - V_\text{eff}''(\phi_\text{cl})\right)\delta(x - x')\:,
\end{equation}
and thus the fluctuation operator is given by
\begin{equation}
    \label{eq:L_operator}
    \text{L} = \frac{\partial^2}{\partial t^2} - \frac{\partial^2}{\partial x^2}+ \frac{\rho}{M^2} - \mu^2 + 3\lambda\phi^2\:.
\end{equation}
Let \(\Psi_n(t, x)\) be the discrete eigenfunctions of \(\text{L}\), with corresponding eigenvalue \(\lambda_n\). The index \(n\) may or may not be an integer, but will take values in some finite subset of \(\mathbb R\). One should also expect a set of continuum modes, which take purely imaginary values of \(n\). These will not determine the stability of the classical configuration and so will not be considered explicitly (see ref.~\cite{Millington2026}), but both discrete and continuum modes would need to be calculated were we to proceed to construct the Green's function via its spectral representation.

Returning to the eigenvalue equation
\begin{equation}
    \text{L}\Psi_n(t, x) = -\lambda_n\Psi_n(t, x)\:,
\end{equation}
we assume that the eigenfunctions oscillate in time with frequency \(\omega\), i.e., \(\Psi_n(t, x) = e^{-i\omega t}\Psi_n(x)\), so that their spatial dependence satisfies
\begin{equation}
    \left(-\frac{\partial^2}{\partial x^2} + \frac{\rho}{M^2} - \mu^2 -\omega^2 + 3\lambda\phi^2\right)\Psi_n(x) = -\lambda_n\Psi_n(x)\:.
\end{equation}
The eigenfunctions are orthogonal, with
\begin{equation}
    \int_{-\infty}^\infty\text{d}x\,\Psi_n(x)\Psi^*_{n'}(x) = \delta_{nn'}\:.
\end{equation}
If we define the parameter \(E_n = \omega^2 - \lambda_n\) and the function \(U(x) = V''(\phi)\), then the eigenvalue problem can be written in the suggestive form
\begin{equation}
    \left(-\frac{\partial^2}{\partial x^2} + U(x)\right)\Psi_n(x) = E_n\Psi_n(x)\:.
\end{equation}
In other words, the spatial part \(\Psi(x)\) of the eigenfunctions of \(\text{L}\) are simply the eigenfunctions of a Hamiltonian operator \(\hat H = -\partial_x^2 + U(x)\) with eigenvalues $E_n$.

\begin{figure}
    \centering
    \begin{subfigure}{\textwidth}
        \centering
        \includegraphics{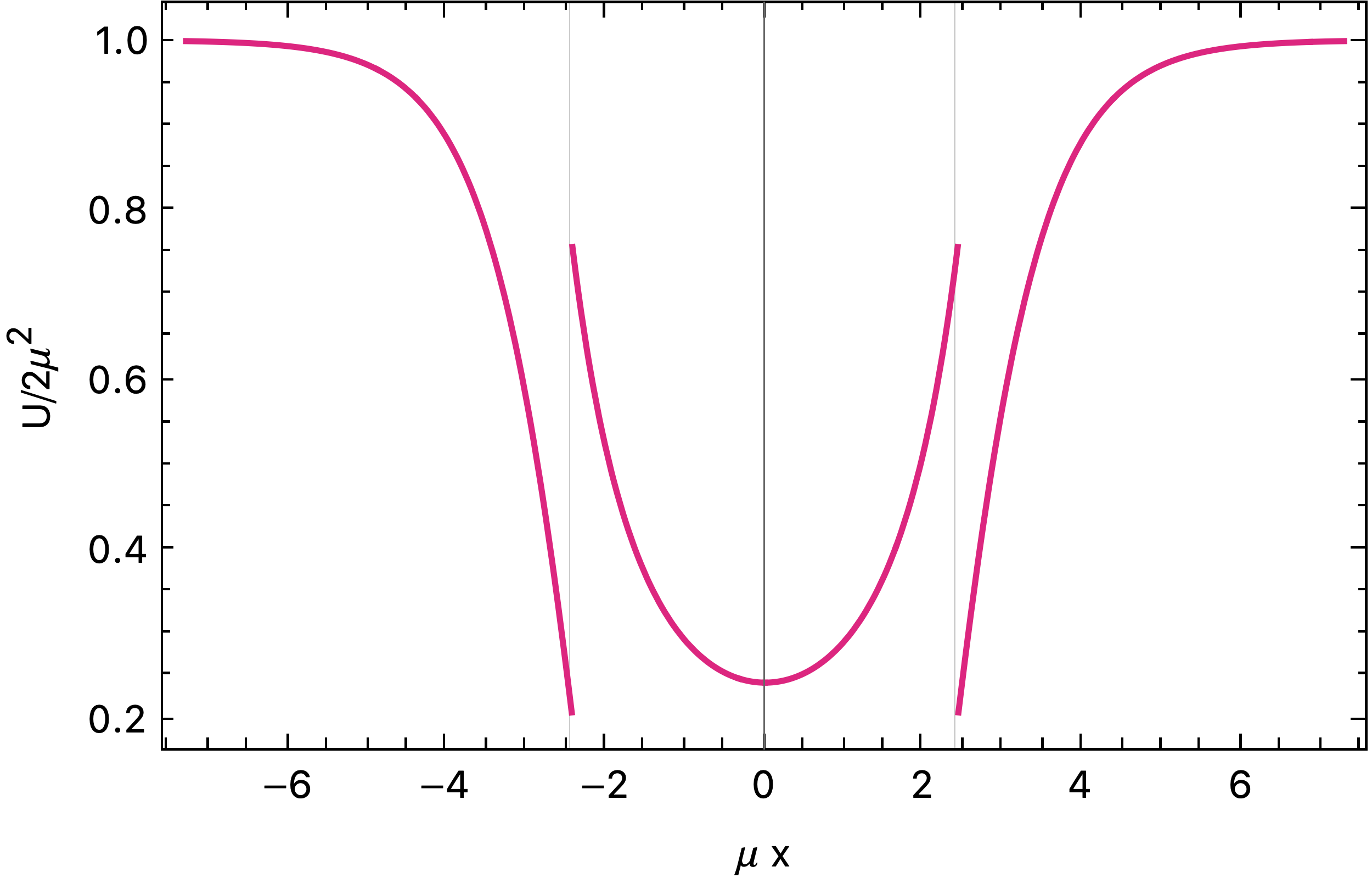}
        \subcaption{The effective potential of the classical configuration as a function of position.}
        \label{fig:eigenpotential_position}
    \end{subfigure}
    \vfill
    \begin{subfigure}{\textwidth}
        \centering
        \includegraphics{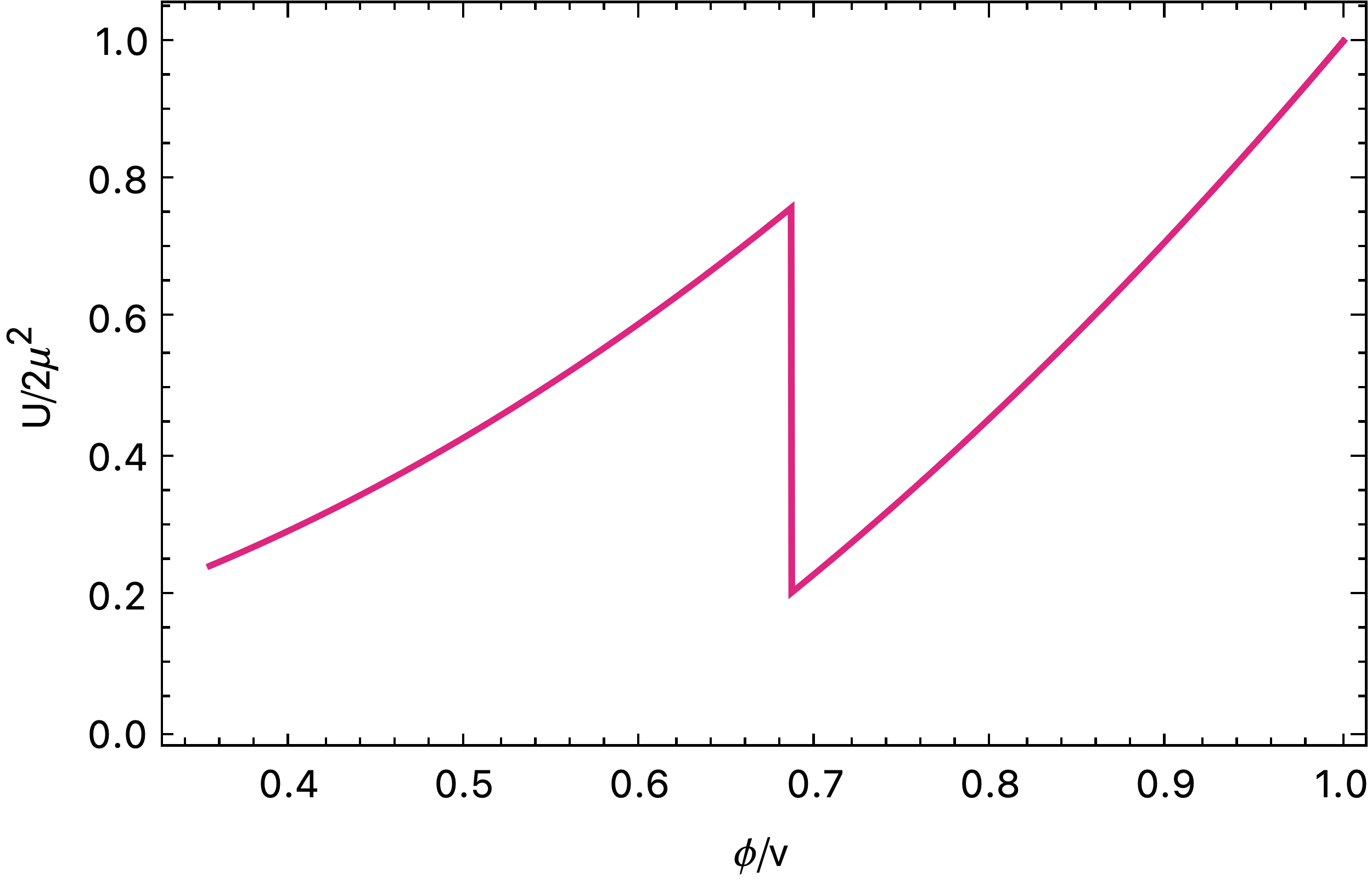}
        \subcaption{The effective potential of the classical configuration as a function of the normalised field. Note that the effective potential is piecewise convex in this representation.}
    \label{fig:eigenpotential_field}
    \end{subfigure}
    \caption{The effective potential of the classical configuration, plotted as a function of position (above) and the classical field (below). The parameter values were chosen for illustrative purposes (\(mu = 2\), \(\rho_0 = 25\), \(M = \sqrt{5}\) and \(R = 1/2\)). The noted features of the effective potential persist across parameter space.}
\end{figure}

For completeness, we catalogue the properties of the potential \(U(x)\). See Figure~\ref{fig:eigenpotential_position} for a graphic representation of the following information, and Figure~\ref{fig:eigenpotential_field} for the potential as a function of the field. For \(x<R\), \(U\) is given by
\begin{equation}
    U(x) = \frac{\rho_0}{M^2} - \mu^2 + 3\mu^2\frac{\phi_0^2}{v^2}\text{nc}^2\left(\xi\mu x,k^2\right)\:.
\end{equation}
As \(x\rightarrow 0\), \(\text{nc}\left(\xi\mu x,k^2\right)\rightarrow 1\). Therefore,
\begin{equation}
    \lim_{x\rightarrow 0}U(x) = \frac{\rho_0}{M^2} - \mu^2 + 3\mu^2\frac{\phi_0^2}{v^2} = \xi^2\mu^2\left(5 - 4k^2\right)\:.
\end{equation}
The potential initially grows slowly, and then rapidly reaches a global maximum
\begin{equation}    
    U_\text{max} = \frac{\rho_0}{M^2} - \mu^2 + 3\mu^2\frac{\phi_0^2}{v^2}\text{nc}^2\left(\xi\mu R,k^2\right)\:,
\end{equation}
before dropping to a global minimum
\begin{equation}
    U_\text{min} = - \mu^2 + 3\mu^2\frac{\phi_0^2}{v^2}\text{nc}^2\left(\xi\mu R,k^2\right)\:.
\end{equation}
For \(x>R\), the potential is
\begin{equation}
    U(x) = -\mu^2 + 3\mu^2\tanh^2\left(\frac{\mu}{\sqrt{2}}\left(x - R\right) + \operatorname{arctanh}\frac{\phi_{R}}{v}\right)\:.
\end{equation}
Since \(\tanh x\rightarrow 1\) as \(x\rightarrow \infty\), we find the lower asymptote is
\begin{equation}
    \lim_{x\rightarrow \infty} U(x) = 2\mu^2\:.
\end{equation}
Therefore, if there are any discrete modes, we should expect them for \(U_\text{min} < E_n < 2\mu^2\). This is essentially the argument for the existence of bound states of the Hamiltonian. Additionally, we should find that the continuous spectrum begins for \(E_n > 2\mu^2\). They correspond to the Hamiltonian's scattering states.

Since the Schrödinger potential is even, one can always find a basis of solutions with definite parity. Thus, once again, we need only solve the problem in the positive domain. Moreover, the eigenmode with the lowest energy is always even, with zero nodes, and the eigenmode with the next lowest energy is always odd, with one node which is necessarily located at the origin. As we climb the energy ladder, the parity of the eigenfunctions oscillates.

The exterior modes \(\Psi^>_{n}\) satisfy
\begin{equation}
    \label{eq:ext}
    \left(\frac{\text{d}^2}{\text{d}x^2} + \mu^2 + E_{n} - 3\lambda\left(\phi_>\right)^2\right)\Psi^>_{n} = 0\:.
\end{equation}
By changing the independent variable to the normalised external field,
\begin{equation}
    \label{eq:u}
    u = \frac{\phi_>}{v} = \tanh\left(\frac{\mu}{\sqrt{2}} x + \dots\right)\Rightarrow \frac{\text{d}}{\text{d}x} = \frac{\mu}{\sqrt{2}}\left(1-u^2\right)\frac{\text{d}}{\text{d}u}\:,
\end{equation}
we transform eq.~\ref{eq:ext} to
\begin{equation}
    \label{eq:legendre}
    \frac{\text{d}}{\text{d}u}\left(\left(1-u^2\right)\frac{\text{d}\Psi^>_{n}}{\text{d}u}\right) + \left(l(l+1) - \frac{n^2}{1-u^2}\right)\Psi^>_{n} = 0\:,
\end{equation}
and recover the general Legendre equation with degree \(l = 2\) and order \(n\) satisfying
\begin{equation}
    \label{eq:order}
    n^2 = 4 - \frac{2 E_n}{\mu^2}\:.
\end{equation}
By similar arguments employed in ref.~\cite{Millington2026}, we may choose to represent the solution set in terms of the Legendre polynomials \(P_2^n\) and \(P_2^{-n}\) and, without loss of generality, take \(n\) to be positive. Then, up to a normalisation factor, the exterior modes are given by
\begin{equation}
    \Psi^>_{n}(x) = A^>_nP_2^{-n}\left(u(x)\right)\:,
\end{equation}
where the \(A_{n}\in\mathbb R\) will be determined later. Rearranging eq.~\ref{eq:order} gives the spectrum
\begin{equation}
    \label{eq:spectrum}
    E_{n} = \frac{\mu^2}{2}(4 - n^2)\:.
\end{equation}
Notice that higher \(n\) means lower energy \(E_n\). We should expect the lowest-lying even eigenmode to have some \(n = n_0\) greater than \(n = n_1\) for the lowest-lying odd eigenmode. If we find an \(n\) which exceeds 2, we have a negative mode along with an instability. If we find an \(n\) equal to unity, we will have a zero mode.

The interior eigenvalue problem ought to give rise to the same spectrum, so we substitute eq.~\ref{eq:spectrum} into the interior eigenvalue equation. The result is
\begin{equation}
    \label{eq:int}
    \left[\frac{\text{d}^2}{\text{d}x^2} - \frac{n^2\mu^2}{2} + 3\mu^2 - \frac{\rho_0}{M^2} - 3\lambda(\phi_<)^2\right]\Psi^<_{n} = 0\:.
\end{equation}
We could change variables to the field profile normalised with respect to \(\phi_0\), i.e., \(u = \text{nc}\left(\xi\mu x, k^2\right)\), but it turns out the substitution
\begin{equation}
    \label{eq:squared_sub}
    z = \text{nc}^2\left(\xi\mu x, k^2\right) \Rightarrow \frac{\text{d}}{\text{d}x} = 2\xi\mu\left(1 - k^2\right)^{1/2}\left(z(z-1)(z - a)\right)^{1/2}\frac{\text{d}}{\text{d}z}
\end{equation}
is more useful, where we have defined
\begin{equation}
    a = \frac{k^2}{k^2 - 1}\:.
\end{equation}
After making this substitution, one will eventually find that the eigenvalue problem may be written
\begin{equation}
    \left[\frac{\text{d}^2}{\text{d}z^2} + \left(\frac{\gamma}{z} + \frac{\delta}{z - 1} + \frac{\epsilon}{z - a}\right)\frac{\text{d}}{\text{d}z} + \frac{\alpha\beta z - q}{z(z - 1)\left(z - a\right)}\right]\Psi^<_{n} = 0\:,
\end{equation}
where \(\gamma = \delta = \epsilon = 1/2\), \(\alpha = -1\), \(\beta = 3/2\) and
\begin{equation}
    q_n = \frac{1}{4(1 - k^2)}\left(\frac{n^2 - 4}{2\xi^2} + \nu^2 - 1\right)\:.
\end{equation}
This is a resolution of the Heun equation \cite{ronveaux_1995}, and the details of its derivation and properties are given in Appendix~\ref{app:heun}. In general, one finds singular points at \(z = 0, 1, a, \infty\), but the only singularity which occurs for us is \(z=1\). There exist two solutions valid in a circle around \(z=1\) which excludes the nearest singularity, which would be \(z=0\). Since \(z\geq 1\), our solutions should be valid in the entire domain.
It is worth noting that, since \(\gamma = \delta = \epsilon = 1/2\), we could also represent the eigenvalue problem as a Lam{\'e} equation (see Appendix~\ref{app:lame}). In principle, either representation is equivalent. In practice, given the wider literature, Heun functions are much easier to deal with.

Our canonical solutions are
\begin{equation}
    \label{eq:heun1}
    H_1(z;n) = H\mathcal\ell\left(1-a, \alpha\beta-q_n,\alpha,\beta,\delta,\gamma;1-z\right)
\end{equation}
and 
\begin{align}
    \label{eq:heun2}
    \nonumber H_2(z;n) =&\, (1-z)^{1-\delta}\\
    &\times H\mathcal\ell\left(1-a,((1-a)\gamma+\epsilon)(1-\delta) + \alpha\beta - q_n;\alpha+1-\delta,\beta+1-\delta,2-\delta,\gamma;1-z\right)\:.
\end{align}
The general solution will be some linear combination of the two. However, definite parity implies that the odd/even eigenfunction will comprise one or the other. Since \(H_1\) does not vanish at the origin, it will contribute to the even eigenfunction. Since \(H_2\) does vanish at the origin, it will contribute to the odd eigenfunction. Let \(\nu\) be a non-negative integer which indexes the eigenmodes. Then, for \(\nu = 0\) and \(\nu\) even, we have
\begin{equation}
    \Psi_{n_{\nu}}(x) = \Theta(x - R)A_{n_{\nu}} P_2^{-n_{\nu}}\left(u(x)\right) + \Theta(R - x)B_{n_{\nu}} H_1(z(x);n_{\nu})\:, 
\end{equation}
and for \(\nu\) odd we have
\begin{equation}
    \Psi_{n_{\nu}}(x) = \Theta(x - R) A_{n_{\nu}}P_2^{-n_{\nu}}\left(u(x)\right) + \Theta(R - x)B_{n_{\nu}} H_2(z(x);n_{\nu})\:,
\end{equation}
where \(\nu\) is a non-negative integer.

\begin{figure}
    \centering
    \begin{subfigure}{0.45\textwidth}
        \includegraphics[width=\textwidth]{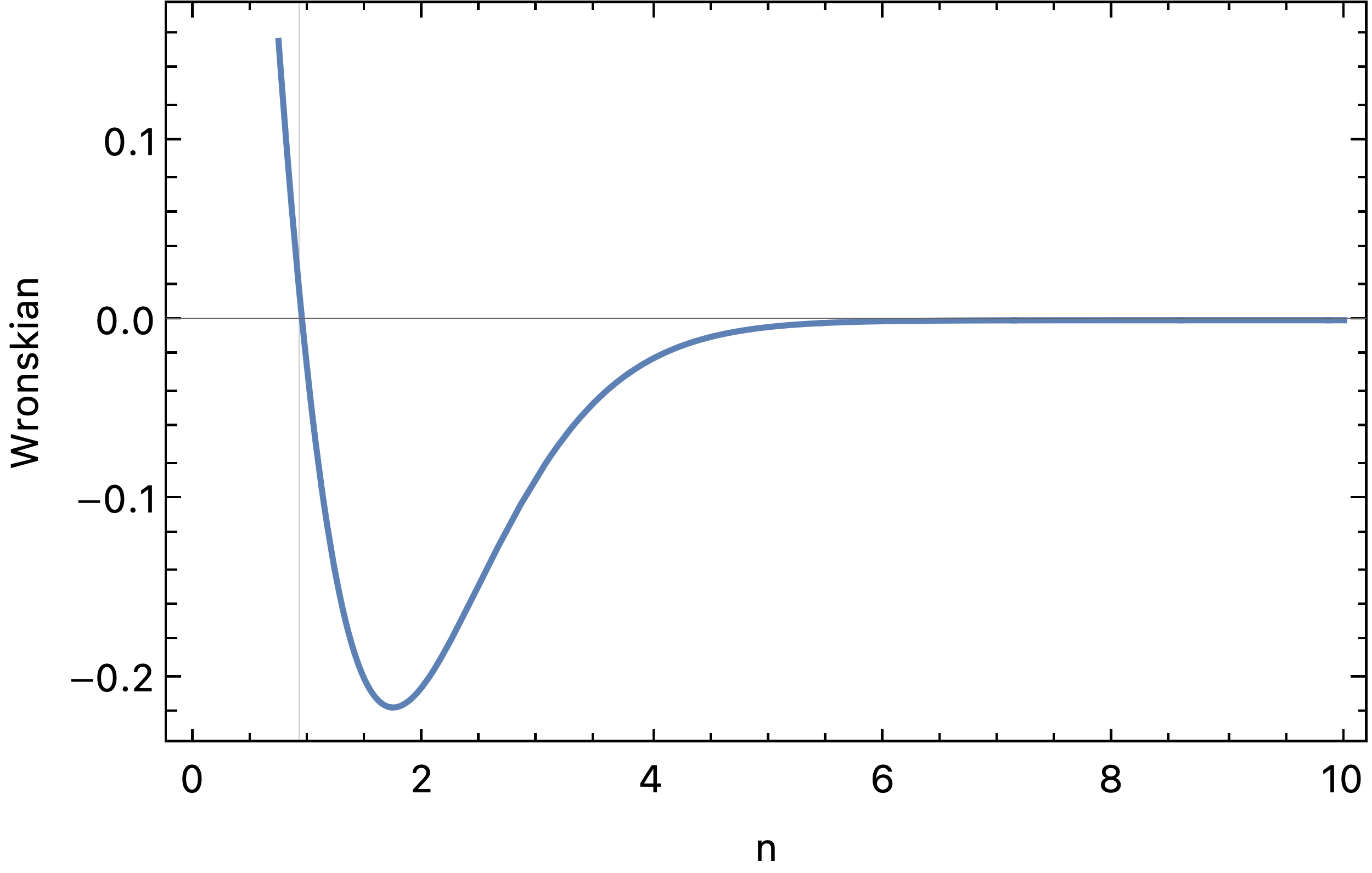}
        \caption{Wro{\'n}skian for the even eigenfunction. The vertical line is at the root \(n = 0.94307\dots\).}
        \label{fig:1d_wronskian_even}
    \end{subfigure}
    \hfill
    \begin{subfigure}{0.45\textwidth}
        \includegraphics[width=\textwidth]{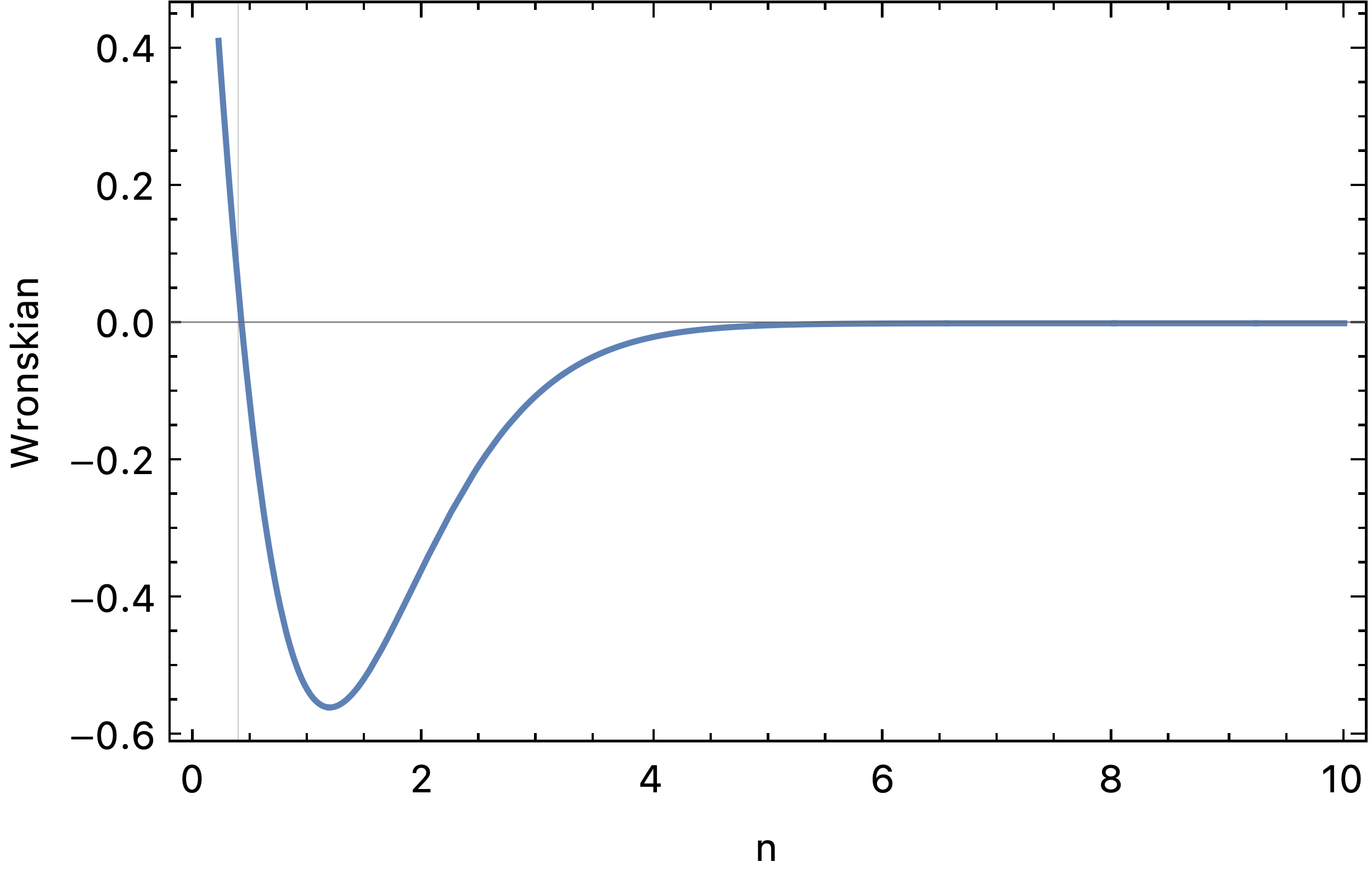}
        \caption{Wro{\'n}skian for the odd eigenfunction. The vertical line is at the root \(n = 0.41732\dots\).}
        \label{fig:1d_wronskian_odd}
    \end{subfigure}
    \caption{A graphical depiction of the roots of the quantisation condition eq.~\ref{eq:1d_quant}.}
    \label{fig:1d_wronskian}
\end{figure}

\begin{figure}
    \centering
    \begin{subfigure}{0.45\textwidth}
        \includegraphics[width=\textwidth]{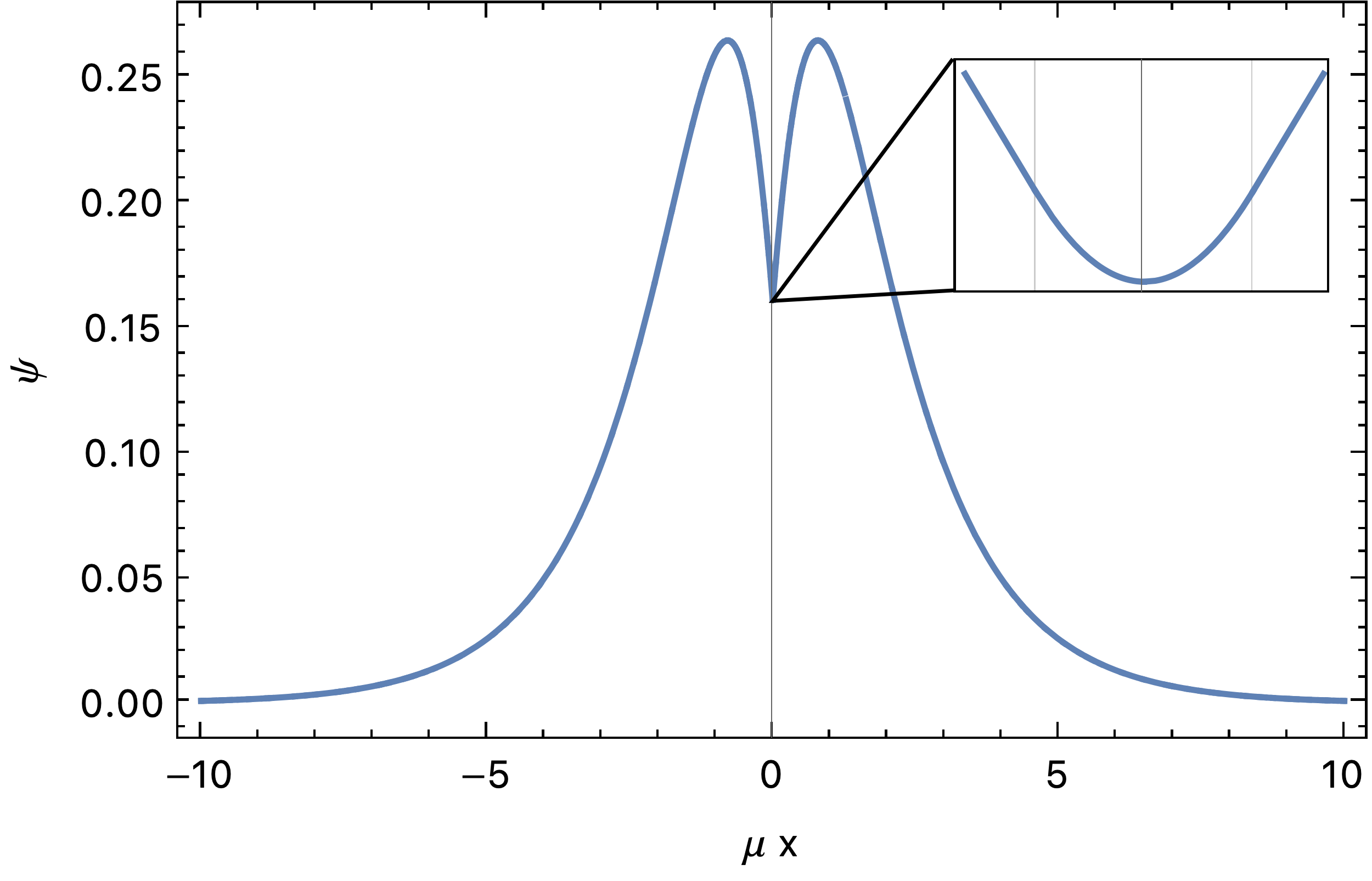}
        \caption{\(\nu=0\)}
        \label{fig:k0}
    \end{subfigure}
    \hfill
    \begin{subfigure}{0.45\textwidth}
        \includegraphics[width=\textwidth]{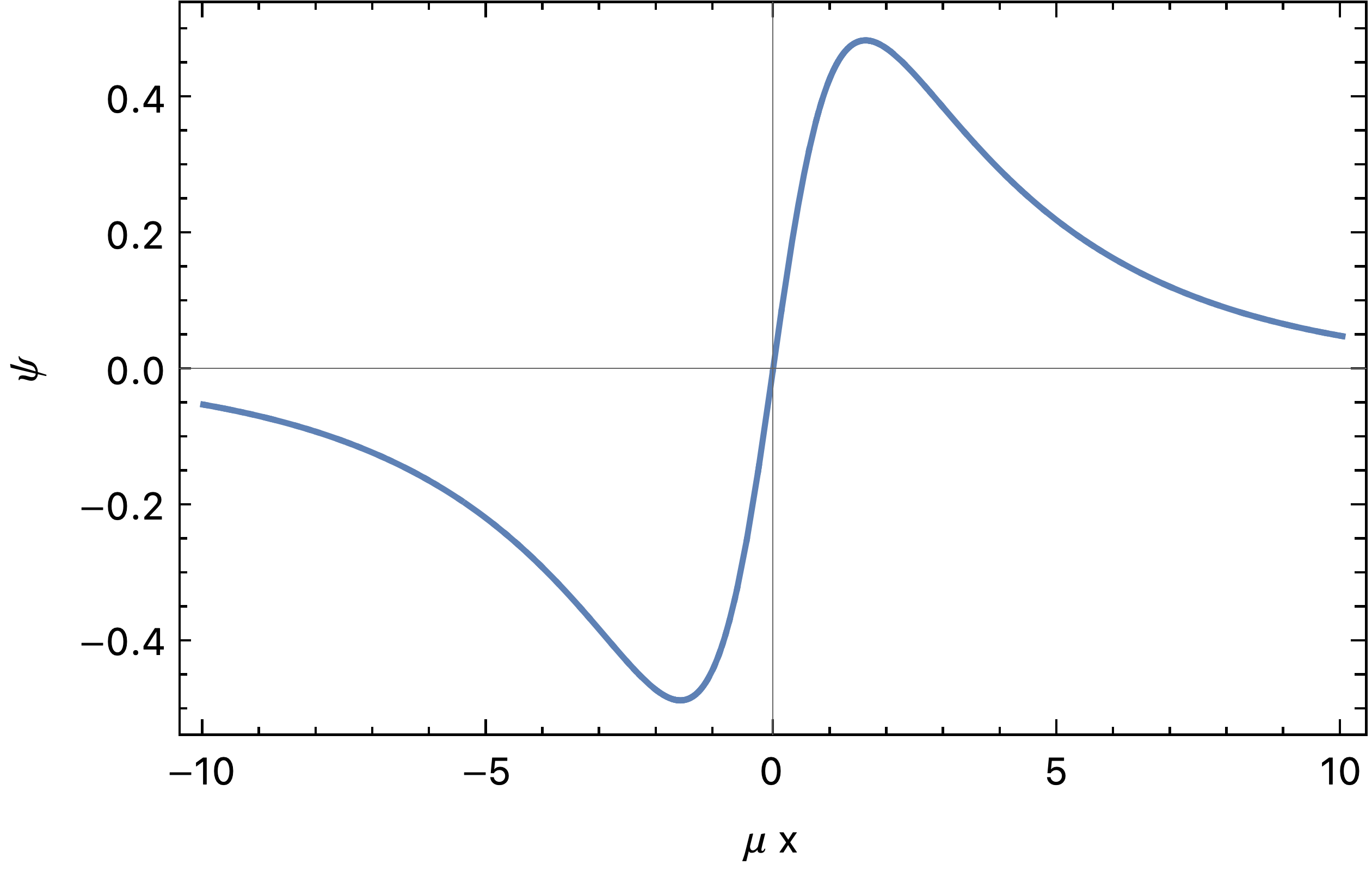}
        \caption{\(\nu=1\)}
        \label{fig:k1}
    \end{subfigure}
    \caption{The discrete eigenmodes of the fluctuation operator.}
    \label{fig:eigenmodes}
\end{figure}

Simultaneously satisfying continuity and differentiability at the surface of the source is equivalent to the condition
\begin{equation}
    \label{eq:1d_quant}
    \left.W\left(H_{1,2}(z(x);n), P_2^{-n}(u(x))\right)\right\vert_{x = R} = 0\:.
\end{equation}
Obviously, this condition cannot be solved in closed form, so we must solve it numerically. Plots of the Wro{\'n}skian at \(x = R\), as a function of \(n\), are given in Figure~\ref{fig:1d_wronskian_even} and Figure~\ref{fig:1d_wronskian_odd}, respectively, for the even and one odd eigenfunction. The roots characterise the complete discrete spectrum, with one even and odd eigenmode. We use Mathematica's inbuilt root-finding method to obtain the solutions. Crucially, all eigenenergies are positive. See Figure~\ref{fig:eigenmodes} for plots of the even and odd eigenfunctions.

\section{Green's Function}
\label{sec_4}
Referring back to eq.~\ref{eq:q_correction}, the next step in our calculation is to determine the Green's function \(G(x, y)\) in the background of the classical field configuration $\phi_\text{cl}$ from Section~\ref{sec_2}. In this section, we will summarise the computation of this Green's function. The methods involved are broadly similar to those presented in our previous work \cite{Millington2026}. The technical details are presented in Appendix \ref{app:technical_green}.

We begin with the equation
\begin{equation}
    \label{eq:green}
    \text{L}G(\mathbf x, \mathbf x';t, t) = - \delta(\mathbf x - \mathbf x')\delta(t - t')\:,
\end{equation}
where \(\text{L}\) is the fluctuation operator (second functional derivative of the action of the classical configuration) given in eq.~\ref{eq:L_operator}. In contrast to the spherically symmetric case, we can take advantage of translational symmetry in the \(y\) and \(z\) direction. 
To that end, we express the Green's function as
\begin{equation}
    G(\mathbf x, \mathbf x';t, t) = \int\frac{\text{d}E\text{d}k_y\text{d}k_z}{\left(2\pi\right)^3}\,e^{-iE(t - t')}e^{i\left[k_y(y-y') + k_z(z-z')\right]}\,G(x, x'; E, \mathbf k)\:,
\end{equation}
where \(\mathbf k = (k_y, k_z)\) is the momentum conjugate to \((y, z)\). After making this substitution, the only coordinate dependence we need to solve for is in the \(x\) direction,
\begin{equation}
    \label{eq:axial}
    \left(-\frac{\partial^2}{\partial x^2} - E^2 + \mathbf k^2 + \frac{\rho}{M^2} - \mu^2 + 3\lambda\phi^2\right)G(x, x';E, \mathbf k) = -\delta(x - x')\:.
\end{equation}

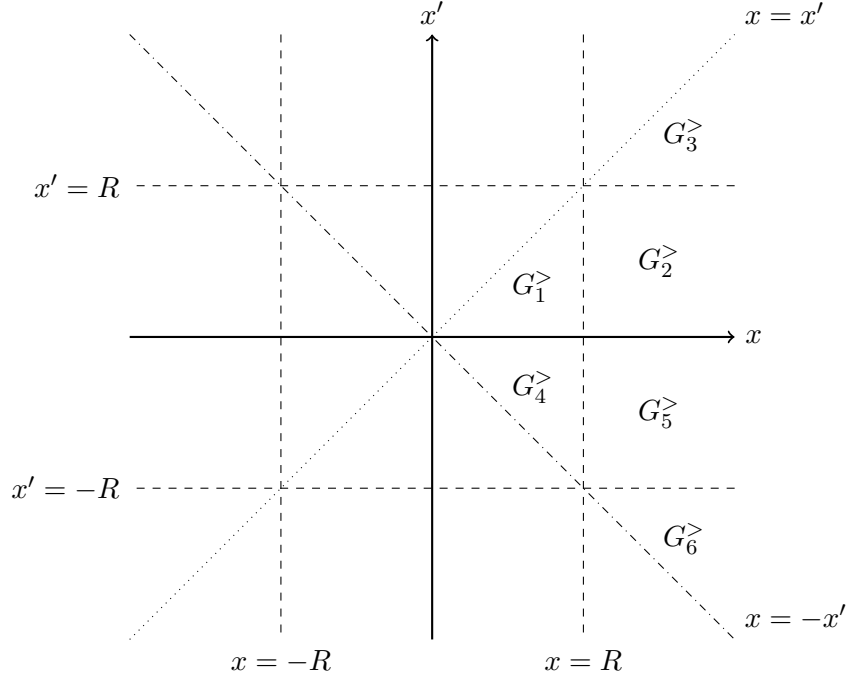
\begin{figure}
    \centering
    \begin{tikzpicture}
    \draw[thick,->] (-4,0) -- (4,0) node[right] {\(x\)};
    \draw[thick,->] (0,-4) -- (0,4) node[above] {\(x'\)};

    \draw[dashed] (2,4) -- (2,-4)  node[below] {\(x = R\)};
    \draw[dashed] (-2,4) -- (-2,-4) node[below] {\(x = -R\)};

    \draw[dashed] (4,2) -- (-4,2) node[left] {\(x' = R\)};
    \draw[dashed] (4,-2) -- (-4,-2)  node[left] {\(x' = -R\)};

    \draw[dotted] (-4,-4) -- (4,4) node[above right] {\(x = x'\)};

    \draw[dashdotted] (-4,4) -- (4,-4) node[above right] {\(x = -x'\)};

    \draw (1.333, 0.666) node {\(G^>_1\)};
    \draw (3, 1) node {\(G^>_2\)};
    \draw (3.333, 2.666) node {\(G^>_3\)};

    \draw (1.333, -0.666) node{\(G^>_4\)};
    \draw (3, -1) node{\(G^>_5\)};
    \draw (3.333, -2.666) node{\(G^>_6\)};

    \end{tikzpicture}
    \caption{A graphical depiction of lines of discontinuity for the Green's function problem.}
    \label{fig:green_domains}
\end{figure}

To solve this problem, we must account for five lines of discontinuity. The Green's function is, of course, continuous, with a continuous first derivative everywhere except where \(x=x'\). Then there are the lines \(x = \pm R\) and \(x'=\pm R\), along which the fluctuation operator is discontinuous. The discontinuity of the delta function is also a line of symmetry for the Green's function, represented by the reciprocity condition \(G^>(x, x'; E, \mathbf k) = G^<(x', x; E, \mathbf k)\). Furthermore, the line \(x = -x'\) is a line of symmetry, since the fluctuation operator is even in \(x\). This amounts to what we shall call the reflection condition \(G(x, x'; E, \mathbf k) = G(-x', -x; E, \mathbf k)\), and tells us how to generate the contributions that lie in the unmarked domains (see Figure \ref{fig:green_domains}).

Each \(G_j^>\) is a product of two functions, one depending on \(x\) and the other depending on \(x'\). Furthermore, these functions are themselves linear combinations of solutions to the eigenfunction equations. The most general forms of the solutions which satisfy the homogeneous boundary condition are as follows,
\begin{align}
    G_1^>(x, x'; E, \mathbf k) &= \frac{1}{W_1}\left(H_1(x) + A_1H_2(x)\right)\left(H_1(x') + B_1H_2(x')\right)\:,\\
    G_2^>(x, x'; E, \mathbf k) &= \frac{1}{W_2}P_2^{-n}(x)\left(H_1(x') + A_2H_2(x')\right)\:,\\
    G_3^>(x, x'; E, \mathbf k) &= \frac{1}{W_3}P_2^{-n}(x)\left(P_2^{n}(x') + A_3P_2^{-n}(x')\right)\:,\\
    G_4^>(x, x'; E, \mathbf k) &= \frac{1}{W_4}\left(H_1(x) + A_4H_2(x)\right)\left(H_1(x') + B_4H_2(x')\right)\:,\\
    G_5^>(x, x'; E, \mathbf k) &= \frac{1}{W_5}P_2^{-n}(x)\left(H_1(x') + A_5H_2(x')\right)\:,\\
    G_6^>(x, x'; E, \mathbf k) &= \frac{1}{W_6}P_2^{-n}(x)\left(P_2^{-n}(x') + A_6P_2^{n}(x')\right)\:.
\end{align}
The Legendre order parameter \(n\) is now
\begin{equation}
    n = \frac{\sqrt{2}}{\mu}\left(\mathbf k^2 - E^2 + 2\mu^2\right)^{1/2}\:,
\end{equation}
as would be anticipated from the planar limit of the spherical case (see ref.~\cite{Millington2026}).

For brevity, we shall define the symbol \(\mathcal W_j^{\operatorname{sgn}(n)}(y)\) by
\begin{equation}
    \mathcal W_j^{\pm}(y) := \lim_{x\rightarrow y^+}W(H_j(x), P_2^{\pm n}(x))\:.
\end{equation}
In terms of this symbol, \(G^>_{1, 4}\) may be written
\begin{align}
    \label{eq:g1}
    G^>_1(x, x'; E, \mathbf k) &= -\frac{1}{2i\xi\mu}\frac{\mathcal W_2^-(R)}{\mathcal W_1^-(R)}\left(H_1(x) - \frac{\mathcal W_1^-(R)}{\mathcal W_2^-(R)}H_2(x)\right)\left(H_1(x') + \frac{\mathcal W_1^-(R)}{\mathcal W_2^-(R)}H_2(x')\right)\;,\\
    \label{eq:g4}
    G^>_4(x, x'; E, \mathbf k) &= -\frac{1}{2i\xi\mu}\frac{\mathcal W_2^-(R)}{\mathcal W_1^-(R)}\left(H_1(x) - \frac{\mathcal W_1^-(R)}{\mathcal W_2^-(R)}H_2(x)\right)\left(H_1(x') - \frac{\mathcal W_1^-(R)}{\mathcal W_2^-(R)}H_2(x')\right)\:,
\end{align}
the functions \(G_{2, 5}^>\) take the form
\begin{align}
    \label{eq:g2}
    G_2^>(x, x'; E, \mathbf k) &= \frac{1}{2\mathcal W_1^-(R)}P_2^{-n}(x)\left(H_1(x') + \frac{\mathcal W^-_1(R)}{\mathcal W^-_2(R)}H_2(x')\right)\:,\\
    \label{eq:g5}
    G_5^>(x, x'; E, \mathbf k) &= \frac{1}{2\mathcal W_1^-(R)}P_2^{-n}(x)\left(H_1(x') - \frac{\mathcal W^-_1(R)}{\mathcal W^-_2(R)}H_2(x')\right)\:,
\end{align}
\(G^>_3\) is
\begin{equation}
    \label{eq:g3}
    G_3^>(x, x'; E, \mathbf k) = -\frac{\pi}{\mu\sqrt{2}}\csc(n\pi)P_2^{-n}(x)\left(P_2^{n}(x') -\frac{1}{2}\left(\frac{\mathcal W_1^+}{\mathcal W_1^-} + \frac{\mathcal W_2^+}{\mathcal W_2^-}\right)P_2^{-n}(x')\right)\:,
\end{equation}
and \(G_6^>\) is given by
\begin{equation}
    \label{eq:g6}
    G^>_6(x, x'; E, \mathbf k) = \frac{i\xi\mu}{2\mathcal W_1^-\mathcal W_2^-}P_2^{-n}(x)P_2^{-n}(x')\:.
\end{equation}
Together, the results of this section can be pieced together in the various domains to give the full Green's function.
\section{Tadpole Contribution and Quantum Correction}
\label{sec_5}
The final step is to compute the renormalised tadpole $\Pi$ appearing in eq.~\ref{eq:q_correction} . This requires us to evaluate the coincidence limit of the Green's function. This quantity is divergent, and we regularise the loop integral with a UV cutoff \(\Lambda\).

Expressing the integral over \(\mathbf k\) in plane polar coordinates with radial coordinate \(k\) yields
\begin{equation}
    G(x) = 2\int_0^\infty\frac{\text{d}E}{2\pi}\int_0^\infty\frac{\text{d}k}{2\pi}\,k\,G(x;E, k)\:,
\end{equation}
where $G(x;E,k)\equiv G(x,x;E,k)$. We next perform a Wick rotation, which will be equivalent to replacing every \(E^2\) with \(-E^2\) and adding a factor of \(i\). We then introduce \(p^2 = E^2 + k^2\), which allows us to write the regularised coincident limit of the Green's function as
\begin{equation}
    G(x) = \frac{i}{2\pi^2}\int_0^\Lambda\text{d}p\,p^2\,G(x; p)\:.
\end{equation}

Now comes the task of determining, as much as possible, an analytic expression for the regularised coincident limit. By the symmetry arguments we have been using throughout the planar analysis, it is sufficient to work in the positive \(x\) domain. We start, as always, exterior to the source, where the coincident Green's function takes the form
\begin{equation}
    G_+(x;E, k) = -\frac{\pi}{\sqrt{2}\mu}\csc(n\pi)P_2^{-n}(x)\left(P_2^n(x)-\frac{1}{2}\left(\frac{\mathcal W_1^+}{\mathcal W_1^-} + \frac{\mathcal W_2^+}{\mathcal W_2^-}\right)P_2^{-n}(x)\right)\:.
\end{equation}
It is possible to express this function analytically in terms of a UV-finite and UV-divergent part. The divergent contribution is contained within the cross term
\begin{equation}
    G_+^{(1)}(x;E, k) = -\frac{\pi}{\sqrt{2}\mu}\csc(n\pi)P_2^{-n}(x)P_2^n(x)\:.
\end{equation}
This is, up to a factor of \(R^2\), the same contribution as was obtained in the thin-wall approximation in ref.~\cite{Millington2026}. Thus, we already know it will lead to the expression
\begin{equation}
    G_+^{(1)}(x) = -\frac{i\mu^2}{16 \pi^2}\left(\frac{2\Lambda ^2}{\mu^2}+2- \left(1-3 u^2\right) \log \left(\frac{\mu^2 }{2\Lambda^2 }\right)-\pi  \sqrt{3} u^2\left(1-u^2\right)\right)\:,
\end{equation}
as expected.

Things are not as straightforward for the interior contribution. It takes the form
\begin{equation}
    G_-(x;E, k) = -\frac{1}{2i\xi\mu}\frac{\mathcal W_2^-(R)}{\mathcal W_1^+(R)}\left[H_1(x)^2 - \left(\frac{\mathcal W_1^+(R)}{\mathcal W_2^-(R)}H_2(x)\right)^2\right]\:.
\end{equation}
Notice that the value of the field at the boundary of the source explicitly appears in every term. The boundary value of the field does not appear in the UV-divergent contribution to the coincident Green's function, and so we found no way to analytically separate the UV-finite part from the UV-divergent part. Of course, it was always the case that the rest of the coincident Green's function would not admit an exact representation, so the result is some loss of accuracy in the numerical method.

Regarding the computational estimate, the renormalisation of the tadpole will be performed numerically. It must be the case that the UV divergence that we find exterior to the source is also present in the interior. As a result, we can determine the counterterms in vacuum, and these are (see refs.~\cite{Garbrecht:2015oea, Millington2026})
\begin{equation}
    \delta \lambda = -\frac{9\lambda^2 }{16 \pi ^2}\left(\ln \left(\frac{\mu^2}{2\Lambda ^2}\right)+5 \right)
\end{equation}
and
\begin{equation}
    \delta m^2 = -\frac{3 \lambda\mu^2}{16 \pi ^2}\left(\frac{2\Lambda ^2}{\mu^2}-\log \left(\frac{\mu ^2}{2\Lambda ^2}\right)-31 \right)\:.
\end{equation}
Instead of using these expressions directly, we numerically renormalise the tadpole contribution via pseudo-counterterms, following ref.~\cite{} (see also ref.~\cite{Udemba:2025csd}). These are the functions \(\Delta m^2(p)\) and \(\Delta\lambda(p)\) that satisfy the integral equations
\begin{equation}
    \frac{1}{2\pi^2}\int_0^\Lambda\text{d}p\,p^2\Delta m^2(p) = \delta m^2\:,
\end{equation}
and
\begin{equation}
    \frac{1}{2\pi^2}\int_0^\Lambda\text{d}p\,p^2\Delta \lambda(p) = \delta\lambda\:,
\end{equation}
(see also ref.~\cite{PhysRevD.92.125022}) which allow us to write the tadpole contribution in the form
\begin{equation}
    \Pi^R(\phi)= \frac{1}{2\pi^2}\int_0^\Lambda\text{d}p\,p^2\left[-3\lambda G(x;p) + \Delta m^2(p) + \Delta\lambda(p)\phi^2\right]\:.
\end{equation}

The only things we must change in our numerical method are the functions we integrate over. The lack of efficient numerical implementations of Heun functions presents a limitation. For areas of the parameter space or experimental setups which have \(\rho_0 \gg \mu^2 M^2\) or \(R\gg \mu^{-1}\), the squared Jacobi elliptic function \(z = \operatorname{nc}\left(\xi\mu x, k^2\right)^2\), which is the argument of the Heun function \(H_{1, 2}(z)\), can range from very small to very large values. More generally, this occurs whenever the setup is such that \(\chi_0 \ll 1\). The current fastest Python implementation \cite{Birkandan2021} (which we have since improved upon slightly \cite{mu_github}) loses accuracy at an unacceptable rate for even moderately sized domains. Furthermore, Mathematica's implementation is impractically slow for large domains. The result is that we cannot speak as generally as we would like, restricting our attention to the few areas of parameter space where convergence can be achieved in Mathematica in a reasonable timeframe.

\begin{figure}
    \centering
    \includegraphics{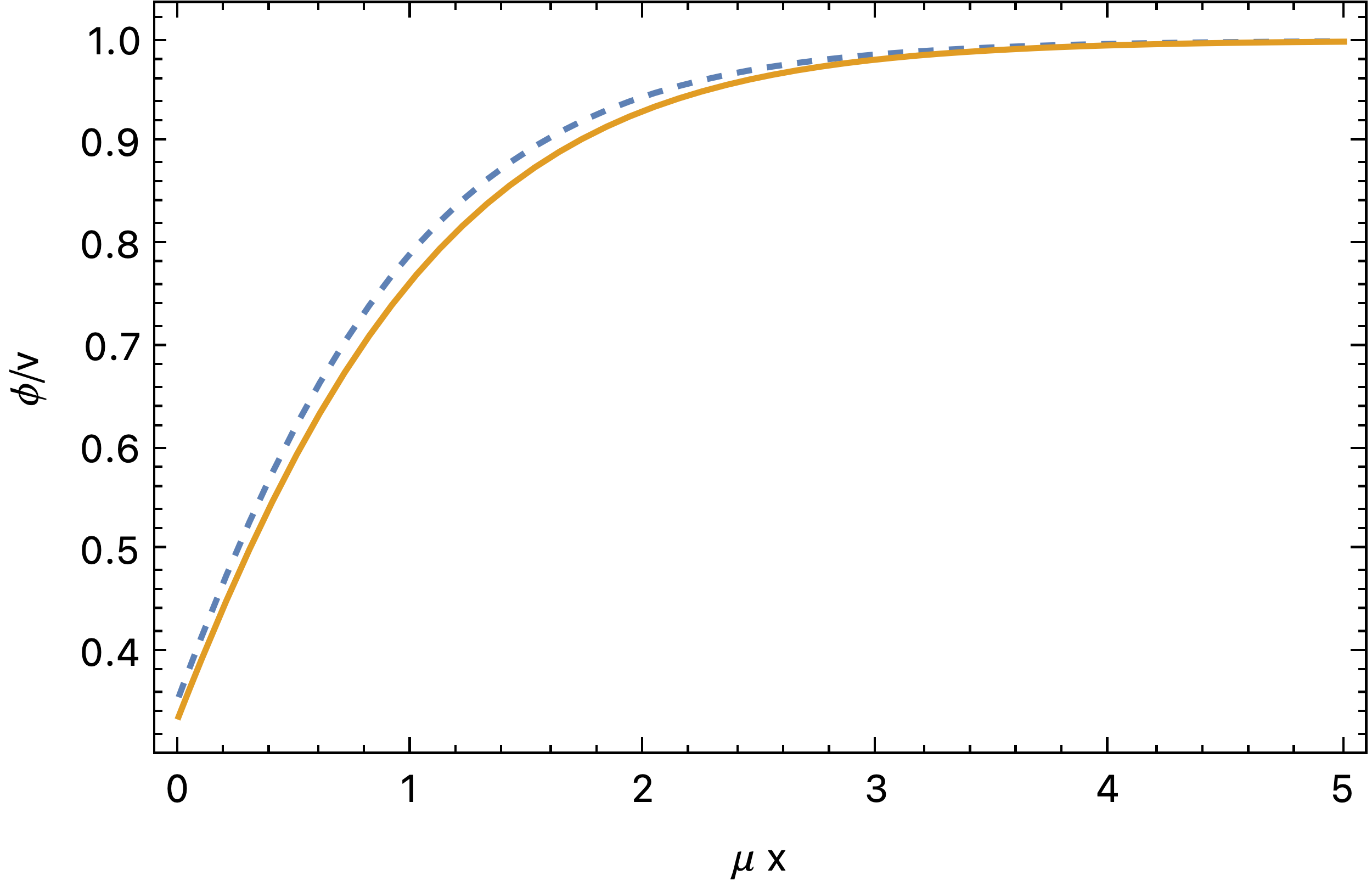}
    \caption{The one-loop (orange, solid) and tree-level (blue, dashed) symmetron field profiles for a planar source. Parameter values were chosen to correspond to CANNEX: \(\rho_0 = \rho_{\text{SiO}_2}\), \(R = 3\,\)mm, \(M = 10^{-3}\,\)GeV and \(\mu = 10^{-1}\,\)eV.}
    \label{fig:planar-clvsqu}
\end{figure}

The setup of CANNEX comprises a 6 mm-thick silicon dioxide source plate \cite{Almasi2015, Sedmik2021}. We obtain the field profile in Figure~\ref{fig:planar-clvsqu} for such a plane, with a mass term \(\mu = 10^{-1}\,\)eV, mass scale \(M = 10^{-3}\,\)GeV and self-coupling \(\lambda = 0.9\) towards the end of the perturbative regime. Much of the correction is due to a change in the slop of the field profile and not a shift in the VEV at one-loop.

\begin{figure}
    \centering
    \includegraphics[scale=1.1]{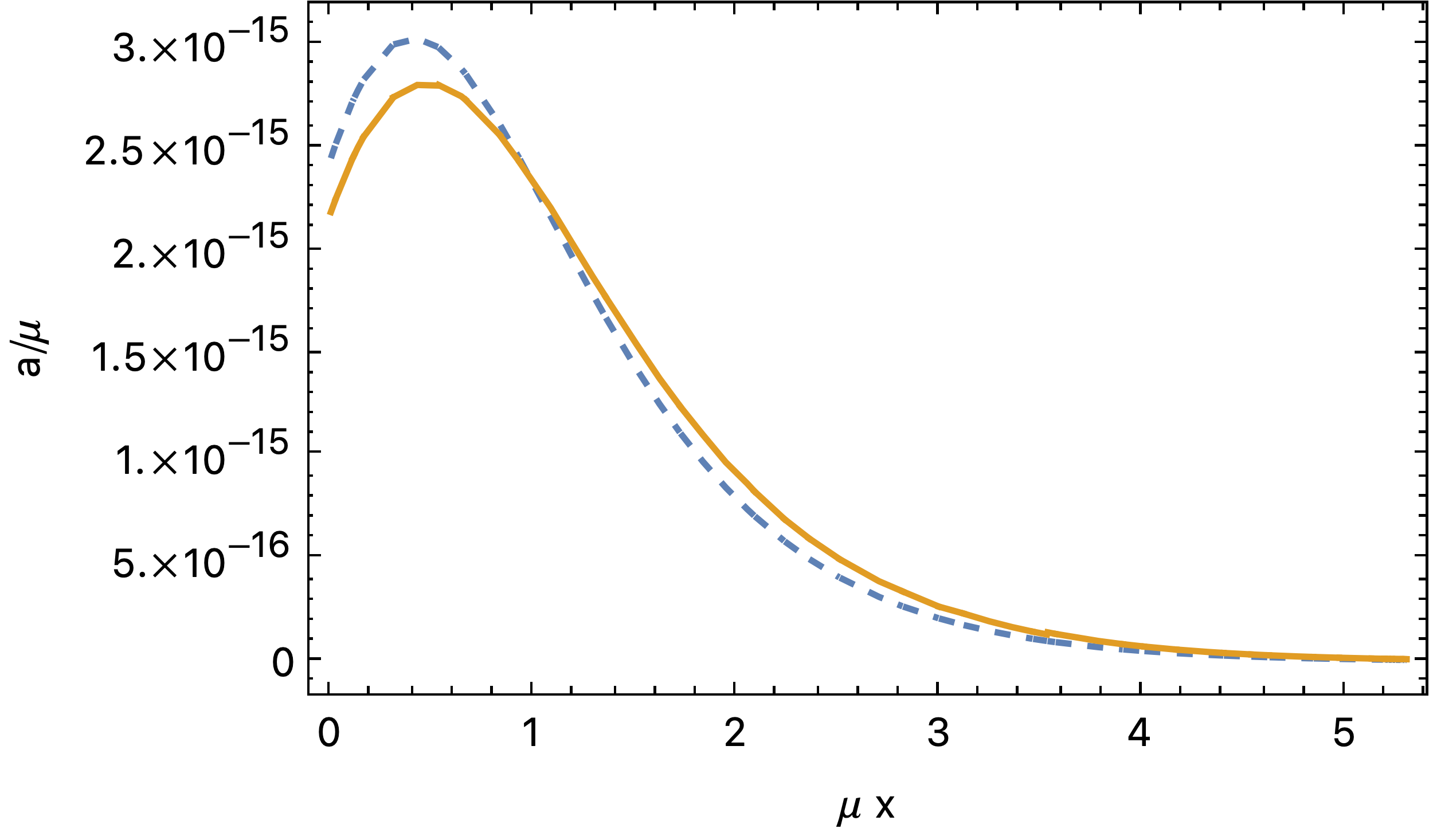}
    \caption{The variation of the one-loop (orange, solid) and tree-level (blue, dashed) symmetron force for a planar source. Parameter values were chosen to correspond to CANNEX: \(\rho_0 = \rho_{\text{SiO}_2}\), \(R = 3\,\)mm, \(M = 10^{-3}\,\)GeV and \(\mu = 10^{-1}\,\)eV.}
    \label{fig:planar-quantum_force}
\end{figure}

A plot of the force per unit mass, or acceleration, is given in Figure~\ref{fig:planar-quantum_force}. Here there is some similarity with the setup presented in ref.~\cite{Millington2026}, and we find broadly consistent results. First, it is apparent that the one-loop correction to the fifth force is more than an overall scaling. Second, the quantum corrections lead to a change in the shape of the fifth-force profile. At a range appropriate for CANNEX, which has a plate separation of between 3 and 30\(\,\mu\)m, we observe a one-loop force which is just over 10\% weaker than the classical, tree-level prediction. At around one Compton wavelength from the source, far out of the detection range of tabletop experiments, the one-loop force then becomes stronger than its classical counterpart. This effect may have interesting implications for the viability of this and similar fifth-force models. Specifically, there may exist a region of parameter space where the symmetron force is further suppressed on length scales comparable to the Solar System, further evading experimental and observational bounds, while being enhanced over galactic distances. Moreover, this change in the fifth force profile will have implications for the optimisation of tabletop experiments aiming to maximise the sensitivity to such forces \cite{Banks:2025vvz, Briddon:2023ayq}.

\section{Conclusions}
\label{sec_6}
This paper has presented a calculation of the leading quantum corrections to the symmetron field profile and fifth force around a planar source of a thickness and density comparable to the source plate of CANNEX. To simplify the analytical calculations, we considered a source of sufficiently large spatial extent to make boundary effects negligible, and obtained a closed form for the classical field configuration in agreement with expressions previously obtained in the literature \cite{PhysRevD.99.024045, PhysRevD.97.064015,PhysRevD.103.084013}. We then performed an analysis of the discrete spectrum of the second functional derivative of the action of this classical configuration, in preparation for the calculation of the Green's function in the following section and to provide a novel probe of the stability of the classical solution. Finally, we numerically renormalised the one-loop correction and obtained estimates for the one-loop field profile and fifth force.

Similar to our previous work \cite{Millington2026}, we have found that the one-loop quantum correction weakens the strength of the fifth force at length scales which are relevant for tabletop experiments. For the chosen parameter region and source properties, this was a 10\% effect. We also reproduced a strengthening of the fifth force around a few Compton wavelengths from the surface of the source, though such distances would be far outside the reach of terrestrial experiments given the value of the mass parameter \(\mu\). Even so, the change in the fifth-force profile induced by these first quantum corrections, and the resulting relative weakening and strengthening of the fifth force as we move away from the source may have important phenomenological relevance to the viability of these classes of scalar-tensor theories. Moreover, we have seen evidence that the inclusion of quantum corrections has implications for the optimisation of source geometries for tabletop fifth-force searches.

The semi-analytic calculations presented in this paper are another step towards a better understanding of how one-loop corrections change the previously classical predictions of symmetron cosmology. Along with our previous work~\cite{Millington2026}, the present results will provide key benchmarks for comprehensive numerical calculations of quantum corrected fifth-force profiles for more physical and/or experimentally relevant source geometries. We leave this and a systematic reassessment of existing constraints to future work.

\section{Acknowledgements}
The authors thank Richard Battye and Christopher McCabe for helpful discussions. This work was supported by the University of Manchester, the Science and Technology Facilities Council (STFC) [Grant No.ST/X00077X/1], and a United Kingdom Research and Innovation (UKRI) Future Leaders Fellowship [Grant No. MR/V021974/2].

\appendix
\section{Lam{\'e} and Heun Functions}
In this appendix, we provide details of our derivation of the Heun equation from the interior eigenfunction equation eq.~\ref{eq:int}. Later on in this section, we will also demonstrate that the equation of motion for the interior eigenmodes may also be represented by Lam{\'e} functions, although their properties make them less flexible for our purposes.
\subsection{Heun Functions}
\label{app:heun}
To determine how our eigenvalue equation transforms, it will be useful to write the coordinate transformation as
\begin{equation}
    \frac{\text{d}}{\text{d}x} = A\sqrt{f(z)}\frac{\text{d}}{\text{d}z}\Longrightarrow \frac{\text{d}^2}{\text{d}x^2} = A^2\left(f(z)\frac{\text{d}^2}{\text{d}z^2} + \frac{1}{2}f'(z)\frac{\text{d}}{\text{d}z}\right)\:,
\end{equation}
where \(A = 2\xi\mu\left(1 - k^2\right)^{1/2}\) and
\begin{equation}
    f(z) = z(z-1)(z - a)\:.
\end{equation}
Since we'll end up dividing though by the factor in front of the second order derivative in \(z\), it makes sense to  consider the expression
\begin{align}
    \nonumber \frac{\text{d}^2}{\text{d}z^2} + \frac{1}{2}\frac{f'(z)}{f(z)}\frac{\text{d}}{\text{d}z} &= \frac{\text{d}^2}{\text{d}z^2} + \frac{z  (z - 1) + (z - 1)(z - a) + z(z - a)}{2z(z - 1)(z - a)}\frac{\text{d}}{\text{d}z}\\
    &= \frac{\text{d}^2}{\text{d}z^2} + \left(\frac{\gamma}{z} + \frac{\delta}{z - 1} + \frac{\epsilon}{z - a}\right)\frac{\text{d}}{\text{d}z}\:,
\end{align}
where \(\gamma = \delta = \epsilon = 1/2\). Next we consider the rest of the equation. For now, we'll just consider its transformation after dividing by \(A^2\),
\begin{align}
    \nonumber \frac{1}{A^2}\left[- \frac{n^2\mu^2}{2} + 3\mu^2 - \frac{\rho_0}{M^2} - 3\lambda\phi(0)^2z\right] &= \frac{1}{4\xi^2\mu^2(1 - k^2)}\left[- \frac{n^2\mu^2}{2} + 3\mu^2 - \frac{\rho_0}{M^2} - 3\frac{\mu^2}{v^2}\phi(0)^2z\right]\\
    \nonumber &= \frac{1}{2(1 - k^2)}\left[\frac{1}{2\xi^2}\left(\frac{4 - n^2}{2} + 1 - \frac{\rho_0}{\mu^2M^2}\right) - 3\frac{\phi(0)^2}{2\xi^2v^2}z\right]\\
    \nonumber &= \frac{1}{2(1 - k^2)}\left[\frac{1}{2}\left(\frac{4 - n^2}{2\xi^2} + 1 - \nu^2\right) - 3(1 - k^2)z\right]\\
    &= \alpha\beta z - q\:,
\end{align}
where \(\alpha\beta = -3/2\) and 
\begin{equation}
    q = \frac{1}{4(1 - k^2)}\left(\frac{n^2 - 4}{2\xi^2} + \nu^2 - 1\right)\:.
\end{equation}

The interior eigenfunctions of the symmetron field for a planar source are Heun functions. These are solutions to Heun's differential equation, which is given by \cite{ronveaux_1995}
\begin{equation}
    \label{eq:heun_app}
    \frac{\text{d}^2f}{\text{d}z^2} + \left(\frac{\gamma}{z} + \frac{\delta}{z - 1} + \frac{\epsilon}{z - a}\right)\frac{\text{d}f}{\text{d}z} + \frac{\alpha\beta z - q}{z(z-1)(z-a)}f(z) = 0\:.
\end{equation}
The parameters are such that the singularity parameter \(a\) is neither \(0\) nor \(1\), and \(\gamma + \delta + \epsilon = \alpha + \beta + 1\). The accessory parameter \(q\) is normally free but, in our case, it depends on \(n\). In the planar case, Heun functions play the same role as the hypergeometric functions of the spherical-planar case. As such, the two functions bear a strong relation \cite{andrews_1999}.

The Fuchs--Frobenius solution around \(z = 0\) has the expansion
\begin{equation}
    f(z) = H\ell(a, q;\alpha, \beta, \gamma, \delta; z) = \sum_{n=0}^\infty c_n z^n\:,
\end{equation}
where the coefficients \(c_n\) are defined by \(c_0 = 1\),
\begin{equation}
    c_1 = \frac{qc_0}{a\gamma}\:,
\end{equation}
and
\begin{equation}
    c_{n+1} = \frac{(n((n-1+\gamma)(1+a) + a\delta + \epsilon) + q)c_n - (n-1+\alpha)(n-1+\beta)c_{n-1}}{a(n+1)(n+\gamma)}\:.
\end{equation}
The linearly independent solution is given by
\begin{equation}
    f(z) = z^{1-\gamma}H\ell\left(a, (a\delta + \epsilon)(1-\gamma) + q;\alpha + 1 - \gamma;\beta + 1 -\gamma, 2-\gamma, \delta;z\right)\:.
\end{equation}
For our purposes, it turns out that the solutions centred around \(z = 1\) are more relevant. The expansions for those functions are given by
\begin{equation}
    f(z) = H\ell\left(1-a, \alpha\beta - 1;\alpha, \beta,\delta,\gamma;1-z\right)
\end{equation}
and 
\begin{equation}
    f(z) = (1-z)^{1-\delta}H\ell\left(1-a,((1-a)\gamma + \epsilon)(1-\delta) + \alpha\beta - q;\alpha + 1-\delta, \beta + 1-\delta, 2-\delta,\gamma;1-z\right)\:.
\end{equation}

Functions denoted by \(H\ell\) are called \emph{local} Heun functions. Given further restrictions on the parameters, one can find solutions that are analytic at two or three singularities. The former are generally called Heun functions, while the latter are called Heun polynomials. Both are too restrictive for our purposes but find applications in quantum mechanics \cite{Ruby2025} and black hole perturbation theory \cite{Wu2026}.
\subsection{Lam{\'e} Functions}
\label{app:lame}
In case of the planar fluctuation modes, the Heun parameters \(\gamma\), \(\delta\) and \(\epsilon\) were equal to \(1/2\) (consequently, \(\alpha + \beta = 1/2\)). When this happens, Heun's equation reduces to the Lam{\'e} equation \cite{ronveaux_1995}. The most compact form is that which is written in terms of Jacobi elliptic functions. To derive it, we define the order parameter \(\nu\) and eigenvalue \(h\) such that
\begin{equation}
    \alpha = -\frac{1}{2}\nu\:,\,\beta = \frac{1}{2}(\nu + 1)\:,\,q = -\frac{1}{4}ah\:.
\end{equation}
This yields the algebraic form,
\begin{equation}
    \frac{\text{d}^2f}{\text{d}z^2} + \frac{1}{2}\left(\frac{1}{z} + \frac{1}{z - 1} + \frac{1}{z - a}\right)\frac{\text{d}f}{\text{d}z} + \frac{ah - \nu(\nu + 1)z}{4z(z-1)(z-a)}f(z) = 0\:.
\end{equation}
At this point, let \(m = a^{-1}\) play the role of the elliptic modulus (the second argument) of the Jacobi sine function and perform the change of variables
\begin{equation}
    z = \operatorname{sn}^2(\zeta, m)\:.
\end{equation}
Then, with the help of standard formulas for the derivatives of Jacobi elliptic functions, we arrive at the compact form \cite{whittaker_watson_1996}
\begin{equation}
    \frac{\text{d}^2f}{\text{d}\zeta^2} + \left[h - \nu(\nu + 1)m\operatorname{sn}^2(\zeta, m)\right]f(\zeta) = 0\:,
\end{equation}
where \(h\) more clearly plays the role of an eigenvalue. 

Regarding the solutions to the Lam{\'e} equation, much is known in the case where \(\nu\) is an integer and \(h\) takes on so-called characteristic values, of which there are \(2\nu + 1\). Given these facts, the solutions are called Lam{\'e} polynomials \cite{whittaker_watson_1996, arscott_1964}. They have finite series expansions and are \(2K(m)\)- or \(4K(m)\)-periodic (in a real or imaginary sense), where \(K(m) = F(\pi/2, m)\) is the complete elliptic integral of the first kind. Herein lies a problem. The solutions we need for the interior planar eigenmodes have non-characteristic \(h\), for which substantially less is known. By theorems due to Arscott \cite{arscott_1964}, we can show that the characteristic solutions do not correspond to the interior eigenmodes.

Using Arscott's terminology, Lam{\'e} polynomials \emph{of the first species} are given by even and odd polynomials in \(\operatorname{sn}(\zeta, m)\). First, consider the even solution,
\begin{equation}
    f(\zeta) = \sum_{r = 0}^\infty a_{2r}\operatorname{sn}^r(\zeta, m)\:.
\end{equation}
Then
\begin{equation}
    f'(\zeta) = \sum_{r=0}^\infty 2r a_{2r}\operatorname{cn}(\zeta, m)\operatorname{dn}(\zeta, m)\operatorname{sn}^{2r-1}(\zeta, m)\:,
\end{equation}
and
\begin{align}
    \nonumber f''(\zeta) =&\, \sum_{r=0}^\infty 2ra_{2r}\left[-\operatorname{dn}^2(\zeta, m)\operatorname{sn}^{2r}(\zeta, m)\right.\\
    \nonumber &-m\operatorname{cn}^2(\zeta, m)\operatorname{sn}^{2r}(\zeta, m)\\
    \nonumber &+\left.(2r-1)\operatorname{cn}^2(\zeta, m)\operatorname{dn}^2(\zeta, m)\operatorname{sn}^{2r-2}(\zeta, m)\right]\\
    =&\,\sum_{r=0}^\infty 2r a_{2r}\left[m(2r + 1)\operatorname{sn}^{2r +2}(\zeta, m) - 2r(m+1)\operatorname{sn}^{2r}(\zeta, m) + (2r - 1)\operatorname{sn}^{2r - 2}(\zeta, m)\right]\:.
\end{align}
Substituting these series expressions into the Lam{\'e} equation yields
\begin{align}
    \nonumber 0 =&\, \sum_{r=0}^\infty a_{2r}\left[\left(h-4 (m+1) r^2\right) \text{sn}^{2r}(\zeta, m)\right.\\
    \nonumber &+m (2 r-\nu ) (\nu +2 r+1) \text{sn}^{2r + 2}(\zeta, m)\\
    \nonumber &+\left.2 r (2 r-1)\text{sn}^{2 r-2}(\zeta, m)\right]\\
    \nonumber =&\,\sum_{r=0}^\infty a_{2r}\left(h-4 (m+1) r^2\right) \text{sn}^{2r}(\zeta, m)\\
    \nonumber &+\sum_{r=1}^\infty a_{2r - 2}m (2 r - 2 -\nu ) (\nu +2 r - 1) \text{sn}^{2r}(\zeta, m)\\
    \nonumber&+\sum_{r=0}^\infty 2a_{2r+2}(r + 1) (2 r + 1)\text{sn}(\zeta, m)^{2 r}\\
    \nonumber =&\,a_0h + 2a_2 + \sum_{r=1}^\infty\left[a_{2r}\left(h-4 (m+1) r^2\right) + a_{2r - 2}m (2 r - 2 -\nu ) (\nu +2 r - 1)\right.\\
    &+\left.2a_{2r+2}(r + 1) (2 r + 1)\right]\text{sn}^{2r}(\zeta, m)\:.
\end{align}
This implies that the coefficients \(a_{2r}\) satisfy the recurrence relation
\begin{equation}
    a_{2r}\left(h-4 (m+1) r^2\right) + a_{2r - 2}m (2 r - 2 -\nu ) (\nu +2 r - 1) + 2a_{2r+2}(r + 1) (2 r + 1) = 0\:,
\end{equation}
with initial condition \(a_0h + 2a_2 = 0\). Consider the specific case in which \(\nu = 2\). One characteristic choice of \(h\) is that which makes \(a_{4} = 0\), which will in turn make \(a_{2N}\) vanish for all integers \(N > 1\). This yields the condition
\begin{equation}
    a_{2}\left(h-4 (m+1)\right) + 6a_{0}m = 0\:.
\end{equation}
Taken together with the initial condition, this gives us two characteristic values of \(h\),
\begin{equation}
    \label{eq:char_h}
    h = 2\left(1 + m \pm\sqrt{1-m+m^2}\right)
\end{equation}
for two different choices of normalisation,
\begin{equation}
    a_0 = \frac{-a_2\left(1 + m\mp\sqrt{1-m+m^2}\right)}{3m}\:.
\end{equation}
It would be miraculous indeed if the interior modes were given by Lam{\'e} polynomials. If they were, eq.~\ref{eq:char_h} would give us an exact solution to the eigenvalue problem. Following this line of reasoning implies that the Legendre order of the interior mode is
\begin{equation}
    n^2 = 2\xi^2\left[2k^2\left(1 + \frac{k^2-1}{k^2} \pm \sqrt{1-\frac{k^2-1}{k^2}+\left(\frac{k^2-1}{k^2}\right)^2}\right) + 1 - 2k^2\right]+4\:.
\end{equation}
For the parameter values considered, we would find values \(n\) which do not correspond to the discrete eigenfunctions found earlier in the paper. More specifically, none of them yield a Legendre order which satisfies the eigenvalue problem. Therefore, the interior planar fluctuation modes are non-characteristic solutions of Lam{\'e}'s equation. Arscott calls such solutions transcendental Lam{\'e} functions \cite{arscott_1964}. Given that they are non-terminating series of Jacobi elliptic functions, and that every numerical implementation of Lam{\'e} functions are actually implementations of Arscott's Lam{\'e} polynomials, we find it more practical to represent transcendental Lam{\'e} functions by Heun functions.
\section{Green's Function: Technical Details}
\label{app:technical_green}
In this appendix, we give the technical details of the calculation which yields the Green's function defined by eqs.~\ref{eq:g1} to \ref{eq:g6}. Specifically, we will demonstrate how to compute the constants, \(W_i\), \(A_i\) and \(B_j\), where \(i \in \{1, 2, \dots, 6\}\) and \(j \in \{1, 2\}\).   

For an analytical problem, having 14 total unknowns presents a reasonable challenge. It is first worth checking that we have the appropriate number of constraints. To begin, we have what we will call the derivative discontinuity condition, which states that the first derivative of the Green's function has a jump discontinuity on the line \(x = x'\), given by
\begin{equation}
    \left.\frac{\partial}{\partial x}G^>(x, x'; E, \mathbf k)\right\vert_{x=x'} - \left.\frac{\partial}{\partial x}G^<(x, x'; E, \mathbf k)\right\vert_{x = x'} = 1\:.
\end{equation}
This provides us with 4 constraints, applying to \(G^>_j\) for \(j=1, 2, 3, 4\). Next we have continuity and differentiability where the various domains overlap. These conditions will turn out to be degenerate, and they apply to 6 different line segments. The final condition is the reflection condition, so called because it may be represented by the relation \(G^>(x, x') = G^>(-x', -x)\). It will be more useful to recast in terms of derivatives as
\begin{equation}
    \left.\frac{\partial}{\partial x}G^>(x, x'; E, \mathbf k)\right\vert_{(x, x') = (s, -s)} + \left.\frac{\partial}{\partial x'}G^>(x, x'; E, \mathbf k)\right\vert_{(x, x') = (s, -s)} = 0\:.
\end{equation}
This condition applies to \(G^>_j\) for \(j = 1, 4, 5, 6\), providing us with our last 4 constraints. Strictly speaking, we also have a `parity' condition
\begin{equation}
    G^>(x, x') = G^>(-x, -x')\:,
\end{equation}
and so the symmetries of the Green's function in \(x,x'\)-space are elements of the Klein four-group \(\mathbb Z_2\times\mathbb Z_2\). However, it is not difficult to show that the parity condition is equivalent to both the symmetry and derivative discontinuity condition. Conversely, the symmetry condition is equivalent to parity and derivative discontinuity.

Let us first consider \(G^>_1\) and \(G^>_4\). Continuity and differentiability implies that, for all \(x \in (0, R)\), we have
\begin{equation}
    \lim_{x'\rightarrow 0^+}G^>_1(x, x'; E, \mathbf k) = \lim_{x'\rightarrow 0^-}G^>_4(x, x'; E, \mathbf k)
\end{equation}
and 
\begin{equation}
    \lim_{x'\rightarrow 0^+}\frac{\partial}{\partial x'}G^>_1(x, x'; E, \mathbf k) = \lim_{x'\rightarrow 0^-}\frac{\partial}{\partial x'}G^>_4(x, x'; E, \mathbf k)\:.
\end{equation}
Taken together, they imply \(B_4 = -B_1\). Next, imposing the reflection condition on \(G^>_1\) gives us \(B_1 = -A_1\) while for \(G^>_4\) we get \(A_4 = A_1\). Finally, the derivative jump condition implies
\begin{equation}
    W_1 = 2A_1 W(H_1(x), H_2(x)) = 2i\xi\mu A_1 = W_4\:.
\end{equation}
Consequently, \(G^>_1\) and \(G^>_4\) can be chosen to depend on a single parameter,
\begin{align}
    G^>_1(x, x'; E, \mathbf k) &= \frac{1}{2i\xi\mu A_1}\left(H_1(x) + A_1H_2(x)\right)\left(H_1(x') - A_1H_2(x')\right)\;,\\
    G^>_4(x, x'; E, \mathbf k) &= \frac{1}{2i\xi\mu A_1}\left(H_1(x) + A_1H_2(x)\right)\left(H_1(x') + A_1H_2(x')\right)\:.
\end{align}
Indeed, it can be found without specifying the constants in any neighbouring functions. Continuity and differentiability across \(x = R\) implies
\begin{equation}
    A_1 = \left.\frac{W(P_2^{-n}, H_1)}{W(H_2, P_2^{-n})}\right\vert_{x = R} = -\frac{\mathcal W^-_1(R)}{\mathcal W_2^-(R)}\:.
\end{equation}

Like \(G_1^>\) and \(G^>_4\), the functions \(G^>_2\) and \(G^>_5\) form a natural pairing. Continuity and differentiability implies \(A_5 = -A_2\) and \(W_2 = W_5\). The derivative jump condition applies only to \(G^>_2\), and yields
\begin{equation}
    W_2 = 2\mathcal W_1^-(R)\:.
\end{equation}
The reflection condition applies only to \(G^>_5\), and implies
\begin{equation}
    A_2 = \frac{\mathcal W^-_1(R)}{\mathcal W^-_2(R)}\:.
\end{equation}

Since \(G^>_3\) and \(G^>_6\) do not share any arguments, solving for one will not immediately help us solve for the other. The former, however, is fairly easy to determine. The derivative jump condition gives us \(W_3\),
\begin{equation}
    W_3 = W(P_2^2, P_2^{-n}) = -\frac{\sqrt{2}\mu}{\pi}\sin(n\pi)\:,
\end{equation}
and continuity on the line \(x' = R\) give us \(A_3\),
\begin{equation}
    A_3 = -\frac{1}{2}\left(\frac{\mathcal W_1^+}{\mathcal W_1^-} + \frac{\mathcal W_2^+}{\mathcal W_2^-}\right)\:.
\end{equation}

Determining \(G^>_6\) is complicated by the fact that continuity of the function and its first derivative is imposed on the line \(x' = -R\). To simplify things, first note that we could have defined \(G^>_4\) in the domain \((0, R)\times (0, -R)\) instead of its upper half triangle and obtain the same result. The same can be said for \(G^>_6\), meaning its simplest form is actually
\begin{equation}
    G_6^>(x, x'; E, \mathbf k) = \frac{1}{W_6}P_2^{-n}(x)P_2^{-n}(x')\:.
\end{equation}
One could alternatively use the reflection condition to prove that \(A_6 = 0\). Now, to find \(W_6\), we impose continuity of the Green's function and its first derivative between \(G_6^>\) and \(G_5^>\). Together, they imply
\begin{equation}
    W_6 = \frac{2}{i\xi\mu}\mathcal W_1^-\mathcal W_2^-\:.
\end{equation}
By appropriately reflecting and rotating these contributions in \(x,x'\)-space, we completely determine the full Green's function \(G(x, x')\).

\bibliographystyle{JHEP}
\bibliography{references.bib}

\end{document}